**Analysis of the 24-Hour Activity Cycle: An illustration examining the association with cognitive function in the Adult Changes in Thought (ACT) Study**

Yinxiang Wu[1,2], Dori E. Rosenberg[1], Mikael Anne Greenwood-Hickman,[1] Susan M. McCurry[2], Cecile Proust-Lima[3], Jennifer C. Nelson[1], Paul K. Crane[2], Andrea Z. LaCroix,[4] Eric B. Larson,[2] Pamela A. Shaw[1]*

1. Kaiser Permanente Washington Health Research Institute, Seattle, WA, USA
2. University of Washington, Seattle, WA, USA
3. Bordeaux Population Health Research Center, INSERM, Bordeaux, France
4. Herbert Wertheim School of Public Health and Human Longevity Science, University of California, San Diego, CA, USA

***Correspondence**
Pamela A Shaw
Pamela.A.Shaw@kp.org

**Abstract**

The 24-hour activity cycle (24HAC) is a new paradigm for studying activity behaviors in relation to health outcomes. This approach inherently captures the interrelatedness of the daily time spent in physical activity (PA), sedentary behavior (SB), and sleep. We describe three popular approaches for modeling outcome associations with the 24HAC exposure. We apply these approaches to assess an association with a cognitive outcome in a cohort of older adults, discuss statistical challenges, and provide guidance on interpretation and selecting an appropriate approach.

We compare the use of isotemporal substitution, compositional data analysis, and latent profile analysis to analyze 24HAC. We illustrate each method by exploring cross-sectional associations with cognition in 1,034 older adults (Mean age = 77; Age range = 65-100; 55.8% female; 90% White) who were part of the Adult Changes in Thought (ACT) Activity Monitoring (ACT-AM) sub-study. PA and SB were assessed with thigh-worn activPAL accelerometers for 7-days. For each method, we fit a multivariable regression model to examine the cross-sectional association between the 24HAC and CASI item response theory (IRT) score, adjusting for other baseline characteristics. We highlight differences in assumptions and the scientific questions that can be addressed by each approach.

ISM is easiest to apply and interpret; however, the typical ISM model assumes a linear association. CoDA specifies a non-linear relationship with time reallocation, through isometric log-ratio transformations that are more challenging to apply and interpret. LPA can serve as an exploratory data analysis tool to classify individuals into groups with similar time-use patterns. Inference on associations of latent profiles with health outcomes need to account for the uncertainty of the LPA classifications, which is often ignored. Cross-sectional data analyses using the three methods did not suggest that less time spent on SB and more in PA was associated with better cognitive function.

The three standard analytical approaches for 24HAC each have advantages and limitations, and selection of the most appropriate method should be guided by the scientific questions of interest and the applicability of each model's assumptions. Further research is needed into the health implications of the distinct 24HAC patterns identified in this cohort.

# 1. Introduction

Physical inactivity and insufficient sleep are well-known risk factors for Alzheimer's Disease and related dementias (ADRD) (Erickson et al., 2019, Livingston et al., 2020, Xu et al., 2020, Sabia et al., 2021). Sedentary behaviors, including activities performed at low energy expenditures and in a sitting or lying down posture (Tremblay et al., 2017), are less consistently associated with cognition, though high-quality studies are lacking (Olanrewaju et al., 2020). Taken together, sedentary time, physical activity, and sleep are the three overarching classifications of movement behaviors people perform across the 24-hour activity cycle (24HAC) when measured by devices such as accelerometers or inclinometers. Most research has examined the contributions of each of these three behaviors independently overlooking the fact that a change in one of the behaviors necessitates a change (either increase or decrease) in the other behaviors (Rosenberger et al., 2019). Given this, there is a great need for research methods that allow researchers to study the 24HAC as a whole.

Several analytical methods have been utilized in the search for an improved understanding of the 24HAC and its association with health outcomes (Livingston et al., 2020, Rosenberger et al., 2019). Traditional approaches, such as regression, cannot be applied to analyze all components of the 24HAC because of the collinearity of the duration of components, i.e., the time in each activity will necessarily add up to the full 24-hours (Rosenberger et al., 2019). Several methods have been used to analyze the association of 24HAC exposures with health outcomes, including isotemporal substitution models (ISM) (Mekery et al., 2009, Grgic et al., 2018), compositional data analysis (CoDA) (Dumuid et al., 2018b, 2020, Janssen et al., 2020), and latent profile analysis (LPA) (Hagenaars and McCutcheon 2002, Evenson et al., 2017, Ekblom-Bak et al., 2020, von Rosen et al., 2020). Few studies to date have applied these methods to assess the association of the 24HAC to tests of cognition.

The goal of this paper is to describe and compare three approaches for modeling outcome associations with the 24HAC exposure: isotemporal substitution models, compositional data analysis, and latent profile analysis. For each approach, we will discuss the model assumptions, interpretation of the models, and highlight some choices that need to be made at the analysis stage. We describe the scientific question addressable by each approach, as well as discuss limitations and some pitfalls to avoid when applying these methods. To illustrate and compare these methods, we will apply each approach to examine the association of the 24HAC with a test of global cognition in a cohort of older adults who were part of the Adult Changes in Thought Activity Monitoring (ACT-AM) sub-study. In the literature, comparisons between ISM and CoDA have been made previously in particular settings (Dumuid et al., 2018a; Biddle et al., 2018), but a comprehensive comparison of these three approaches has not been previously considered. We emphasize the practical application and interpretation of these three models in relation to the cognitive outcome. We also provide sample code in the R software (R Core Team 2021) on GitHub (*https://github.com/yinxiangwu/24HAC_illustrations*), as well as sample syntax for the Latent GOLD LPA software, so that the presented analyses can be readily adapted to other settings.

This work is organized as follows. We first describe the ACT cohort and study data used to illustrate each method. We then introduce each of the three methods, describing the methodology in detail and illustrating the method with an analysis of the ACT cohort. We further compare and contrast each of the approaches and provide considerations for how to choose the approach that best addresses the scientific question of interest. Finally, we conclude with a brief discussion.

# 2. Study Data

*2.1 Study Population*

The Adult Changes in Thought (ACT) study is a longitudinal cohort study of older adults whose aim is to better understand aging, brain aging, and dementia (Kukull et al., 2002). ACT enrolls dementia-free individuals over age 65 who were randomly selected from members of the Kaiser Permanente Washington integrated health care system (KPWA, originally Group Health). The original cohort (N=2581) was enrolled between 1994-1996, with additional enrollment 2000-2003 (expansion cohort, N=811), and beginning in 2004 the ACT Study began ongoing enrollment with a goal to maintain at least 2000 at-risk individuals by replacing those who discontinued, died, or developed dementia (Kukull et al., 2002, Gray et al., 2013). Participants are invited to return for evaluation at 2-year intervals for the ultimate purpose of identifying incident cases of dementia. The study procedures were approved by the institutional review boards of Kaiser Permanente Washington (formerly Group Health Cooperative) and the University of Washington, and participants provide written informed consent.

In April 2016, the ACT-Activity Monitor (ACT-AM) sub-study began inviting ACT participants to wear activPAL accelerometers for 7 days following their regular biennial assessment visits (Rosenberg et al., 2020). Persons who were wheelchair dependent, living in a nursing home, receiving hospice or care for another critical illness, or who showed evidence of cognitive problems during the biennial visit were not asked to participate. Persons who consented were instructed how to wear the device and how to complete a daily log that provided information about device use and time in bed. They also completed an additional take-home survey that included questions about self-reported physical activity and sedentary behavior. The device, daily log, and questionnaire were returned to the research team by mail after one week. Between 2016 and 2018, 1,135 participants consented to wear the activPAL. For these analyses, 1,034 participants with at least 4 valid days of activPAL wear data were included.

*2.2 Assessment of the 24-Hour Activity Cycle*

Details of the activity monitoring device protocols for the ACT-AM sub-study were described previously (Rosenberg et al., 2020). Briefly, waking activity was measured with a thigh-worn activPAL3 micro (PAL Technologies, Glasgow, Scotland, UK) worn on the front central thigh of either leg 24-hours/day for approximately 1 week. As a thigh-worn accelerometer, the activPAL detects both movement and posture (i.e., sitting/lying, with the thigh horizontal, vs. standing, with the thigh vertical). Consequently, activPAL classifies behaviors at the 1-second level as either sitting, standing, or stepping.
Daily time in bed was measured as a surrogate for sleep via participant self-report using a paper log to record in-bed and out-of-bed times each day of wear. Daily wake and sleep times were not constrained to the 24-hour day, and instead were defined by a participant's daily in-bed and out-of-bed schedule, which meant the specific length of any given wear day for a participant varied and could have been greater or less than 24-hours. We defined each full day as out-of-bed time to the next day's out-of-bed time. Based on best practices for free-living activity measurement, a minimum of 4 days with 10 or more hours of waking wear time, as defined by the presence of valid device data during participant self-reported waking periods, was required to be included in analyses (Donaldson et al., 2016, Migueles et al., 2017).

We used proprietary PAL Technologies software to extract event-level files. Events files were then processed by collapsing consecutive activities of the same activity type using a batch processing package `activpalProcessing` in the R software (Lyden 2016). Self-reported time in bed, which the device captures as sitting/lying time, was removed from calculation of waking activity metrics. For simplicity, we refer to time in bed as "sleep". Here we defined the waking 24-hour activity cycle as time sedentary (i.e. sitting/lying down), standing (which also includes very light movement), and physical activity. Specifically, we calculated daily measures of total sitting time (minutes/day), total standing time (minutes/day), and total stepping time (minutes/day) were then calculated during the waking hours of each valid day.

*2.3 Cognition Measures*

ACT participants are evaluated using the Cognitive Abilities Screening Instrument (CASI) at study entry and at each biennial follow-up visit. The CASI consists of 40 items that evaluate attention/concentration, orientation, short- and long-term memory, language abilities, visual construction, verbal fluency, and executive functioning (abstract reasoning and judgment) (Teng et al., 1994). Raw CASI scores range from 0 to 100, with higher scores indicating better cognitive performance. For these analyses, the CASI was scored using item response theory (CASI-IRT), which corrects for potential issues related to unequal interval scaling (Crane et al., 2008). CASI-IRT scores were standardized based on the larger ACT cohort enrollment scores of the study sample such that a 1-unit difference in CASI-IRT can be interpreted as approximately a 1 standard deviation (SD) unit difference in cognitive performance (Ehlenbach et al., 2010; Crane et al., 2008).

2.4 *Other Participant Descriptive Measures*
We examined several other participant characteristics. Objective physical function was derived from a composite of three physical performance tasks completed at the ACT-AM enrollment visit: gait speed (average of two 10-ft timed walks), chair stand time (time needed to move from a seated position in a chair to a standing position, repeated five times), and grip strength as measured by handheld dynamometer (average of three attempts in the dominant hand) (Rosenberg et al., 2020). Each task was scored from 0 to 4 points (higher = better) and then summed to create a physical function score ranging from 0 – 12. For these analyses, two score categories (any impairment 0 to 10; no impairment 11 – 12) were included. Ability to walk half a mile was based on a single self-reported yes/no item "Because of health or physical problems, do you have any difficulty walking one-half mile (about 5 or 6 blocks)?" (McCurry et al., 2002) For persons endorsing walking problems, level of difficulty was further differentiated (no difficulty, some difficulty, a lot of difficulty, unable).

In addition to self-reported daily time in bed used in the 24HAC, descriptive measures of participants' typical sleeping patterns were collected via self-report. Self-reported sleep quality (very poor, poor, fair, good, very good) was based upon a single item from the PROMIS 8-item sleep disturbance scale (Buysse et al., 2010, Yu et al., 2012). A summary of average time in bed (< 6 hours, 6 to 9 hours, > 9 hours) was derived from the daily log kept by participants during the week that they wore the accelerometer devices (described above).

Other participant characteristics were collected from the ACT study visit most proximal to the date of device wear, which was typically the first day of device wear. These included: age (< 74 years, 75 to 84 years, 85+ years), sex (male vs. female), race/ethnicity (non-Hispanic White vs. other), education (some college/post-secondary education vs. high school education or less), measured body mass index (BMI, $kg/m^2$), depressive symptoms from the 10-item Center for Epidemiologic Studies Depression Scale (CES-D; Andresen et al., 1994, Mohebbi et al., 2018), and self-rated overall health (Kohout et al., 1993).

2.5 *Descriptive statistics of the cohort*
Table 1 provides a summary of the subject characteristics, including univariate summaries of the time in each activity considered for the 24HAC from 1034 participants with at least 4 valid days of activPAL wear data. The sample had mean age 77 years (standard deviation (SD) = 7 years, range = [65, 100] years), 55.8% were female, 90% were White, 1.4% were Hispanic. 92.1% of the subjects reported good to excellent self-rated health, 74.6% had no difficulty walking half a mile, and the mean (SD) of CASI-IRT score was 0.61 (0.69) SD units. The mean (SD) time spent on each of the four behaviors, sit, stand, step, and sleep, was, 10 (2), 4 (1.6), 1.4 (0.7), and 8.5 (1.1) hours per day. The median [inter-quartile range (IQR)] total time per day is 1440 [1436, 1445] mins per day.

## 3. Statistical Methods and Analysis of the 24HAC

In this section, we describe three analytical approaches: isotemporal substitution models (ISM), compositional data analysis (CoDA), and latent profile analysis (LPA). For each approach, we describe the method and illustrate its application with the analysis of the 24HAC and its cross-sectional relation to the CASI-IRT score in the ACT cohort. We will then summarize key features of each approach and provide guidance on selecting the most appropriate approach based on features in the data and the scientific question of interest.

3.1 Isotemporal substitution

The 24HAC paradigm imposes a statistical challenge that prevents application of linear regression models with the time spent on each behavior entered in the model simultaneously as a covariate, due to collinearity. By dropping the intercept term, it becomes possible to include all behaviors in one model, often called a "partition model," which can be used to estimate the association of each behavior by holding time in all other behaviors constant. However, when total time per day is fixed, it is not sensible to estimate the effect of a specific behavior without considering other behaviors it displaces. To obtain more reasonable interpretation, a mathematically equivalent model called the isotemporal substitution model (ISM) was adopted, which estimates the "substitution effect" associated with reallocating time from one behavior to another (Mekery et al., 2009). ISM has been widely used in physical activity epidemiology. For example, the ISM approach was applied to examine the association of physical activity intensities with physical health and psychosocial well-being in older adults (Buman et al., 2010), and with cardiovascular disease risk biomarkers using the cross-sectional U.S. National Health and Nutrition Examination Survey data (Buman et al., 2014). Mekary et al., 2013 applied the ISM approach with a time to event outcome to physical activity data from the Nurses' Health Study.

Consider an example where each minute of the 24-hour day is classified into one of four activities: sleeping, sitting, standing, and stepping. The ISM is formulated by including the total activity and all but one of the activity variables – the activity you will explore displacing – in the model. For example, with a continuous health outcome an ISM that leaves out the time stepping can be formulated, as below:

$$E(Y) = \beta_0 + \beta_1\, Sit + \beta_2\, Stand + \beta_3\, Sleep + \beta_4\, Total + \gamma^T X \qquad (1)$$

where $E(Y)$ abbreviates the conditional mean of the health outcome given the time allocation variables (*Sit, Stand, Sleep, Total* measured on the same unit, e.g., minutes in a 24-hour day), and other covariates $X$. In our ACT data, the total amount of time per day, captured by *Total*, slightly varies across individuals according to their sleep schedules and is only approximately 24 hours (see Table 1), and hence a separate intercept term can be included in the model without causing perfect collinearity. When *Total* is exactly a constant 24 hours/day for every subject, only one of the intercept or Total terms can be included in the model; however, this is often not necessary if day length varies, say due to using out-of-bed to out-of-bed time (or in-bed to in-bed time) to define a day. The ISM is a standard regression model (e.g. a linear regression model in equation (1)), so it can be easily fit by statistical software. By omitting one behavior, e.g., stepping in model (1), and controlling for the total time a day, there is no longer perfect collinearity and the coefficients can be interpreted as the estimated effect of time reallocation. In particular, $\hat{\beta}_1$ can be interpreted as follows: when comparing two populations with the same values on the covariates, and the same amount of time spent on standing and sleeping, but with one population spending 1 unit of time/day (e.g., 1 minute/day) more in sitting and the same amount of time less in stepping than the other population, the estimated difference in the mean health outcome is $\hat{\beta}_1$ and similarly for $\hat{\beta}_2$ and $\hat{\beta}_3$. Such comparisons are referred to as the "substitution effect". The coefficients for *Total* and the intercept in this model aren't necessarily meaningful. When total time per day is a constant and there is no intercept term, $\hat{\beta}_4$ functions as the intercept, with the interpretation as the estimated mean outcome for a hypothetical population with 0 mean times spent on sitting, standing, and sleeping, all activity as stepping, all other continuous covariates set at 0 and categorical covariates set at their baseline levels. Each of the variables Sit, Stand, Sleep, and Step could be centered at a set of values $(c_1, c_2, c_3, c_4)$, respectively that add to the fixed Total (e.g., 24 hours), so that $\hat{\beta}_4$ is the expected outcome for the profile $(c_1, c_2, c_3, c_4)$. Generally, however, ISMs are designed to answer scientific questions about effects of time reallocations between activity behaviors, without a particular focus on the intercept coefficient. Importantly, the time reallocations implied by the values of the beta coefficients should not be interpreted as causal effects when the model is fit to observational data, particularly when data are cross-sectional as in our illustration below, despite the commonly used language of "time allocation" in the typical application of ISM that seems to imply causality or an actual change in behavior.

The linear model assumption in model (1) implies a constant substitution effect between any two behaviors regardless of baseline value of the displaced behavior and a symmetric result when the substitution is reversed. This assumption is sometimes too stringent and unrealistic. An easy way to explore a potential nonlinear substitution effect is to fit separate ISMs in groups defined by different intensity levels of an activity behavior. For example, when analyzing the effects of reallocating time from stepping to other behaviors, data can be divided into two groups defined by whether mean step time is above or below a meaningful cut-off. Differential effect estimates from ISMs fit to the two groups can signal potential nonlinearity. Alternatively, a more flexible ISM could be fit with each activity term modeled by a spline function, while keeping the total activity as a linear term, as in Foster et al., 2020. The flexible ISM still enjoys the interpretation of substitution effects but avoids making the linear assumption. The optimal trade-off between smoothness and goodness of fit can be determined by either performing cross validation or minimizing the generalized cross validation (GCV) criteria. The significance of the association for each behavior in the nonlinear ISM model can be tested via a Wald like test (Wood 2006). The nonlinear ISM analysis can be done in R with package "mgcv". Sample R code for this analysis is provided on GitHub (*https://github.com/yinxiangwu/24HAC_illustrations*).

Lastly, through combinations of the regression coefficients, the ISM can also be used to estimate the expected difference in mean outcome between two populations that have different 24-HAC profiles (i.e., different mean time spent on each behavior), but the same values for all other model covariates, which may also be of scientific interest.

*3.1.1 ISM Illustration: Associations between activity behaviors and CASI-IRT in the ACT cohort*

Four linear ISMs adjusted for age, sex, years of education, race/ethnicity, BMI, depressive symptom scores, and self-rated health conditions were fit to the ACT data (excluding 34 (3.3%) with missing covariates), with each of the four activities omitted from the model one at a time. The associations of 30-minute time reallocations between any two types of activity are summarized in Table 2. The ISM model with Step dropped suggested that reallocating 30 minutes/day from sitting, standing, or sleeping to stepping was associated with 0.027 [-0.011, 0.064], 0.033 [-0.011, 0.078], and 0.027 [-0.014, 0.067] SD units higher mean (95% CI) CASI-IRT score, respectively. Symmetric results can be observed for any two activities that would be exchanged. For example, the effect of reallocating 30 minutes from stepping to sitting in this model would be the negative of when the reallocation is reversed (i.e., in the ISM with Sit dropped). This is a direct result of the linear model assumption. In this example, none of the estimated effects of reallocating time mutually between sitting, standing, and sleeping were statistically significant.

Nonlinearity of effects of time reallocation was explored by fitting separate ISMs in groups defined by step time per day with 60 min/day (approximately 1st sample quartile) as a cut-off. No significant associations were found in the subgroups (Table 2). The nonlinear ISM dropping step time was also fit to allow for non-linear associations using penalized cubic splines with 5 equally spaced knots for each activity, maintaining Total time as a linear term, and the model with the smallest GCV value was selected. The effects of substituting each activity on CASI-IRT score in this non-linear model were still nonsignificant with p-values > 0.1.

*3.2* Compositional data analysis

Compositional data analysis (CoDA) is another widely used analytic approach to handle 24HAC data and its associations with health outcomes (Chastin et al., 2015, Dumuid et al., 2018b, Biddle et al., 2018, McGregor et al., 2021). Janssen et al., 2020 conducted a systematic review of CoDA studies examining associations of 24HAC with diverse health outcomes in adults. For CoDA, the fundamental unit of observation is the multivariate vector of the proportions or percentages of the 24 hours that are spent in each type of activity. In the ACT study, we will consider the 24HAC composed of the sleep, sit, stand and step behaviors. One scientific question of interest in the 24HAC paradigm is how different profiles of 24HAC can affect a health outcome. One advantage of CoDA is that it provides a natural way to compare health outcomes between two compositions, including substitution of one behavior for another. For example, given a hypothetical baseline composition, e.g., 10h (41.7%) sitting, 3h (12.5%) standing, 2h (8.3%) stepping, and 9h (37.5%) sleeping, reallocating 2.4h (10%) time per day from sitting to standing corresponds to altering the baseline composition into another composition i.e., 7.6h (31.7%) sitting, 5.4h (22.5%) standing, 2h (8.3%) stepping, and 9h (37.5%) sleeping.

Before illustrating how a regression-based CoDA works, we first introduce a fundamental feature of CoDA called "scale invariance" (Aitchison 1994). Scale invariance means that the total, or the absolute value in a composition, is irrelevant in the analysis and only the relative proportions are of consequence in an outcome model. For example, the composition of sedentary behavior (SB), light-intensity physical activity (LIPA), and moderate-to-vigorous physical activity (MVPA) is often of interest in a physical activity study. The following two activity compositions: 5h (50%) SB, 3h (30%) LIPA, 2h (20%) MVPA and 2.5h (50%) SB, 1.5h (30%) LIPA, 1h (10%) MVPA, will be modeled to have the same mean outcome by CoDA because the two compositions are equivalent. This may not always be a reasonable assumption, particularly when the total activity varies across individuals. For example, in some studies of physical activity (Chastin et al., 2015, Biddle et al., 2018, McGregor et al., 2021), an accelerometry device is worn to capture the percentages of time spent on different types of activities, but the device may not be worn for the whole day. The total time the device is worn will vary by individuals and the sampled composition may not be representative due to selection bias, e.g., people are more likely to wear the device when they are more active. When 24HAC is of interest, the total amount of time per day is the same, or is approximately the same, for most subjects (e.g., in the ACT data, see Table 1), thus, the scale invariance assumption is considered reasonable.

Visualizations and compositional descriptive statistics of 24HAC can be helpful, before fitting any models. Figure 1 displays 24HAC compositions of sit, stand, step, and sleep for our ACT cohort through so-called ternary diagrams, a common tool to visualize composition with 3 parts. Since the 24HAC of interest here consists of four activity behaviors, we plotted four ternary diagrams (Figure 1 A-D), with each graph representing a sub-composition of three activity behaviors. From Figure 1, we can see how sub-compositions are distributed and possibly associated with the outcome. For example, Figure 1(A), shows the 3-part sub-compositions formed by Sit, Stand, and Step. The data points are colored to indicate the CASI-IRT outcome, as an exploratory look at the unadjusted association with this outcome. Most data points are located around the lower left corner of the diagram, indicating that most individuals spend relatively more time sitting than standing and stepping, and there is a tendency for higher CASI-IRT scores (more orange points) to be observed from individuals with higher percentages of stepping time. Similar patterns can be observed from Figure 1(C) and 1(D) where the other sub-compositions involving stepping are considered.

Commonly used compositional descriptive statistics are the compositional mean (center) and variation matrix. The compositional mean can be created by rescaling the vector of geometric means of each behavior, so that the sum of the scaled components equal 100% and the resulting vector is still a composition (Aitchison 1994). Supplemental Appendix A1.1 shows that the compositional mean defined this way has a natural interpretation of the center of a sample of compositions. A variation matrix for the log-ratio (Aitchison 1994) is used to describe the interdependence between every pair of behaviors (Biddle et al., 2018, McGregor et al., 2020, Dumuid et al., 2020). An off-diagonal value close to 0 means the two parts are highly proportional in the observed data. Supplemental Table S1 shows an example of the variation matrix for the sleep, sit, stand, and step 24HAC composition in the ACT study.

CoDA relies on the isometric log-ratio (*ilr*) transformation, which transforms each D-part composition to a unique D-1 vector on a new coordinate system where each new coordinate is a log-ratio which falls along the real line (Egoczue et al., 2003). An example of the *ilr*-transformation is given in equations (2-4), where $z_1, z_2, z_3$ defines the new coordinates for the transformed data; the numbers preceding the log-ratios are normalizing constants, necessary for the desirable mathematical properties of the transformed coordinates (e.g., distance preserving orthonormal basis) (Egocue et al., 2003). See Supplemental Figure S1 for a visualization of this transformation and Supplemental A1.2 for a more detailed description of the general procedure for this transformation.

$$z_1 = \sqrt{\frac{3}{4}} \ln \frac{Sit}{(Stand \times Step \times Sleep)^{1/3}} \quad (2)$$

$$z_2 = \sqrt{\frac{2}{3}} \ln \frac{Stand}{(Step \times Sleep)^{1/2}} \quad (3)$$

$$z_3 = \sqrt{\frac{1}{2}} \ln \frac{Step}{Sleep} \quad (4)$$

The coordinate $z_1$, usually called the *pivot coordinate*, can be interpreted as the log-ratio between the numerator behavior and the geometric mean of the other (denominator) behaviors. A specific set of *ilr*-coordinates can be chosen to capture a particular comparison of geometric means to aid interpretation of that coordinate. Compositional data represented using different *ilr*-coordinates contain essentially the same information and analysis results will be the same regardless of which set of coordinates is used (Pawlowsky et al., 2015). Since the *ilr*-transformation creates a set of continuous variables that are no longer collinear, the transformed compositional data can then be analyzed with standard techniques, e.g., multivariate analysis of variance (James test) or regression analysis.

*3.2.1 Interpreting the CoDA regression model*

We consider how time reallocations between activity behaviors are cross-sectionally associated with CASI-IRT scores in the ACT cohort. It is convenient to create four sets of *ilr*-coordinates with each behavior in turn being singled out as the numerator in the pivot coordinate $z_1$. Four linear regression models were fit with CASI-IRT scores as the outcome, with the resulting *ilr*-coordinates ($z_1, z_2, z_3$) as predictors. Each regression model is adjusted for the other subject characteristics of interest (***X***): age, sex, race/ethnicity, BMI, education level, depressive symptoms, and self-rated health conditions. Once fitted linear regression models are obtained, a little more work is needed to translate the estimated coefficients for those *ilr*-coordinates in a meaningful way in terms of original compositional data.

Considering the following fitted CoDA model,

$$E(Y) = \beta_0 + \beta_1 z_1 + \beta_2 z_2 + \beta_3 z_3 + \gamma^T X \quad (5)$$

where $Y$ is the CASI-IRT score, $z_1$ corresponds to the pivot coordinate $\sqrt{\frac{3}{4}} \ln \frac{Step}{(Sit \times Stand \times Sleep)^{1/3}}$, $z_2 = \sqrt{\frac{2}{3}} \ln \frac{Sit}{(Stand \times Sleep)^{1/2}}$, $z_3 = \sqrt{\frac{1}{2}} \ln \frac{Stand}{Sleep}$ and $X$ is vector of baseline covariates considered.

A direct but not meaningful interpretation of the $z_1$ coefficient is that, holding $z_2$, $z_3$, and other covariates constant, one unit increase in $z_1$ is associated with $\hat{\beta}_1$ increase in the mean outcome. Observing that $z_1$ is the logarithm of the ratio between stepping time and the geometric mean of the time spent in other three behaviors, it is possible to link a difference in $z_1$ to a difference in the ratio. To make more meaningful interpretation in terms of 24HAC composition, we consider differences relative to a referent or baseline composition in order to inform what magnitude difference in $z_1$ is a meaningful difference for the population under study. Suppose the compositional mean calculated over the entire sample is chosen as the baseline (referent) composition, and we are interested in the effect of increasing step by a factor of $(1 + r)$. All other components should simultaneously be decreased by another factor $(1-s)$ to maintain $z2$ and $z3$ constant and ensure all parts sum up to 100%, which leads to the formula $s = r \cdot \frac{x1}{1-x1}$, where $x1$ is the value of the first part in the chosen baseline composition (e.g. step) and $-1 < r < \frac{1-x1}{x1}$ (Dummuid et al., 2018b). For example, the compositional mean in our ACT cohort is 10.23h (42.6%) sit, 3.68h (15.3%) stand, 1.24h (5.2%) step and 8.85h (36.9%) sleep; an increment in stepping time by 10 mins (13% relative to the stepping time in the baseline composition) will require simultaneous decrease in each of the remaining behaviors by 0.7% (i.e. about 4.4 mins sit, 1.6 mins stand, 3.8 mins sleep). By this derivation, the difference compared to the chosen baseline composition is then equivalent to incrementing $z_1$ by $\sqrt{\frac{3}{4}} \log(\frac{1+r}{1-s})$, and keeping $z_2$ and $z_3$ constant; thus, the estimated effect on the outcome is quantified by $\hat{\beta}_1 \cdot \sqrt{\frac{3}{4}} \log(\frac{1+r}{1-s})$. The associated confidence intervals can be obtained by using the variance estimate of $\hat{\beta}_1$. We illustrate this below in our analysis of the ACT Study. The interpretations of the three other possible pivot coordinates can be done similarly, by iteratively changing which component is in the numerator of $z_1$ and repeating this process. Note, the interpretation of $\hat{\beta}_2$ or $\hat{\beta}_3$ alone is not meaningful because it is impossible to increase $z_2$ or $z_3$ while holding the other *ilr*-coordinates constant.

With the fitted CoDA regression model, we can calculate the associated change in the mean outcome between any two given compositions; however, to avoid extrapolation, it is recommended to only consider compositions within the main range of the data. Considering the example raised at the beginning of this section, suppose we want to estimate the difference in the mean outcome between two groups of individuals alike on all covariates, but differing in 24HAC profiles, with one group spending 10h (41.7%) sitting, 3h (12.5%) standing, 2h (8.3%) stepping, and 9h (37.5%) sleeping per day, and the other group spending 7.6h (31.7%) sitting, 5.4h (22.5%) standing, 2h (8.3%) stepping, and 9h (37.5%) sleeping per day. The estimated difference in the mean outcome between the two groups can be quantified as $-0.1\hat{\beta}_1 - 0.48\hat{\beta}_2 + 0.44\hat{\beta}_3$. Interested readers can refer to Supplement A1.3 for more details about this calculation.

In our data analysis, the following R packages were used to conduct CoDA: `ggtern` (Hamilton and Ferry 2018) for ternary diagrams, `robCompositions` for *ilr*-transformations, `Compositional` for testing differences in compositional means between groups, and `deltacomp` for visualizing effects of time reallocations. R code is also provided (*https://github.com/yinxiangwu/24HAC_illustrations*).

*3.2.2 CoDA Illustration: Analysis of 24HAC in the ACT cohort*

In this section, we first present descriptive summaries of the 24HAC for the ACT cohort. We then provide two different applications of the CoDA analysis: one which considers the effect of increasing time in a particular activity, while proportionally decreasing the other activities; and one which considers a composition that captures a pairwise time reallocation. For each of these two examples, the compositional mean was chosen as the reference composition, the calculation for which was described in Section *3.2.1*.

Sitting accounted for the largest proportion (42.6%) of a day, i.e., about 10.2 hours/day, followed by 36.9% (about 8.9 hours/day) taken up by sleeping. Only 15.3% (about 3.7 hours/day) and 5.2% (about 1.2 hours/day) were spent on standing and stepping, respectively. The comparison of the 24HAC compositional means by different participants characteristics is shown in Supplemental Table S2. Female participants spent about 40 mins less per day in sitting but about 30 minutes more time in standing; older participants, especially those aged more than 85 years, spent more time in sitting, but less in both standing and stepping, than those aged 65-74 years; participants with physical function scores greater than 10, or with no difficulty walking half a mile, spent at least 30 mins less time in sitting than those with physical function scores less than 10, or with a lot of difficulty walking half a mile; longer sleepers who spent more than 9 hours/day sleeping tend to be less active in standing and stepping than those who sleep less, whereas few differences were observed by sleep quality. There was a tendency for subjects spending less time in sitting and more time in stepping to have higher CASI-IRT scores, which will be further explored in the multivariable regression below. Due to the large sample size, relatively small differences in the compositions were statistically significant at the level of 0.05, using the James multivariate analysis of variance test on the *ilr*-transformed 24HAC.

Table 3 summarized regression coefficient estimates for the four CoDA pivot coordinates, each of which quantifies the effect of increasing time in one behavior by a factor while simultaneously decreasing time in the other behaviors by another factor. None of these time reallocations were statistically significantly associated with changes in the CASI-IRT scores; however, a higher proportion of time in stepping was weakly associated with higher CASI-IRT scores. This is more evident in Figure 2 which shows predicted differences in mean outcome for reallocating different amounts of time from one behavior to the other behaviors relative to baseline composition (set as the compositional mean in the entire sample). It should be noted that slightly nonlinear and asymmetrical results were observed in the effects of reallocating time from and to stepping. For example, spending 30 mins per day more stepping was associated with 0.026 (95% CI: [-0.006, 0.058]) units higher mean CASI-IRT scores, whereas reversing this time reallocation was associated with 0.038 (95% CI: [-0.085, 0.009]) lower mean CASI-IRT scores. For each of the other behaviors (sitting, standing, and sleeping), reallocating time from or to that behavior was associated with very little change in the CASI-IRT.

Figure 3 presents the effects of reallocating time between any pair of behaviors. Increasing stepping time at the expense of decreased time in either sitting, standing, or sleeping were associated with higher mean CASI-IRT score (Figure 3, 3rd column of graphs). More specifically, reallocating 30 mins per day to stepping from sitting, standing, and sleeping was respectively associated with 0.024 [-0.007, 0.056], 0.031 [-0.011, 0.073], and 0.026 [-0.007, 0.058] higher mean CASI-IRT score, though statistically not significant. No significant effects were observed for time allocations between other pairs of behaviors not involving stepping, with results similar to those from the ISM approach.

*3.3 Latent profile analysis*

Latent profile analysis (LPA) assumes that there is a latent categorical variable that classifies individuals into different subpopulations, thereby identifying groups of individuals with distinct patterns (profiles) of responses for a given set of continuous "indicator" variables. For example, in the 24HAC context, if the three indicator variables were percent of time spent in sedentary behavior, standing, and stepping during the 24-hour day, different profiles may have different mean levels for each of these variables. LPA also generally assumes the indicator variables follow a finite mixture of multivariate normal distributions with each profile having its own specific mean vector and covariance matrix. It is generally assumed that mean vectors differ across profiles, but additional assumptions can be made with respect to whether or not variance matrices of profile indicator variables vary across profiles. Further details are provided in Supplemental Appendix A.2.1. Once the number of latent profiles and variance-covariance structures are fixed, the latent profile model parameters can be estimated using standard statistical software, such as the tidyLPA package in R (Rosenberg et al., 2019), Mplus (Muthén and Muthén 2017) and LatentGold (Vermunt and Magidson, 2021). The fitted model yields the probability of each individual belonging to each of the different profiles (a posterior probability). It is worth mentioning that the parameters are estimated via the expectation-maximization (EM) algorithm (Dempster et al., 1977), an iterative estimation procedure. The EM algorithm requires initial parameter values and can be sensitive to the chosen start values in settings where there may not be a unique solution (solution only a locally, not globally, optimal fit). Thus, running LPA using multiple starting values is recommended until the best solution based on log-likelihood value can be replicated (Berlin et al., 2014). Further details will be discussed in our illustration below.

It is not a straightforward task to decide on the number of latent profiles or the variance-covariance structure. One common strategy is to assume a flexible (i.e., varying by class) variance-covariance structure and fit a series of models with different specifications for the number of classes. A final decision for class number is based on the following commonly used criteria: (1) model fit statistics, including Akaike's Information Criterion (AIC) (Akaike, 1987), Bayesian Information Criteria (BIC) (Schwarz, 1978), sample adjusted Bayesian Information Criteria (SABIC) (Sclove, 1987), and Consistent Akaike's Information Criterion (CAIC) (Bozdogan, 1987); (2) statistical comparison between model fit of $k$ profiles and $k-1$ profiles with p-value < 0.05 favoring the model with $k$ profiles, using Bootstrap likelihood ratio test (BLRT) (McLachlan and Peel 2000) or the Lo-Mendell-Rubin likelihood ratio test (LMR) (Lo, Mendell and Rubin, 2001); (3) statistics that weight model fit, parsimony, and the performance of the classification, e.g. integrated complete likelihood-BIC (ICL-BIC) (Biernacki et al., 2000) (4) profile sample size: although distinct rare groups can exist, a small profile sample size may indicate potential overfitting; and (5) interpretability or clinical utility of resulting profiles. Nylund et al., 2007 performed a simulation study and advocated BIC and BLRT for selecting the correct number of classes. The simulation study in Tein et al., 2013 showed that AIC and entropy were not reliable criteria. Sometimes, subjective judgement based on a mixture of criteria is necessary. For instance, a subjective judgement could be made when deciding between $k$ and $k-1$ profiles, whether the larger number of profiles generated groups that appeared sufficiently distinct, or whether one group appeared too small.

Because of the normality assumption of LPA, it is recommended to always inspect the empirical distributions of profile indicator variables both before and after applying LPA. Although the distribution of a mixture of normal distributions is likely to look nonnormal, which makes it difficult to check the normality assumption at the data pre-processing stage, a highly skewed or heavy tailed distribution often signals a violation of the assumption. In such cases, performing appropriate transformations, e.g., natural logarithm or square root transformation, and/or an outlier analysis is worth considering (Vermunt et al., 2002).

As mentioned before, a key feature of 24HAC data is the co-dependence between activity behaviors. This co-dependence prevents us from applying LPA to all 24HAC variables because it will lead to a degenerate (rank-deficient) covariance matrix. Two possible solutions to consider are (1) apply LPA to the same ilr-transformed variables used in CoDA (see Section 3.2 more details of ilr-transformation) (see Gupta et al., 2020, von Rosen et al., 2020) or (2) drop one activity behavior variable from the LPA (see Jago et al., 2018). Although the first approach retains the full composition, one drawback is the *ilr*-transformation could generate skewed profile indicators unsuitable for LPA. For the second approach, determining which variable to leave out from the LPA model can be made based on scientific considerations, e.g., prior belief of which set of activity behaviors are the driving indicators of underlying profiles or dropping a behavior due to being highly correlated with another behavior.

After obtaining a final model for latent profiles (i.e., the fitted parameters for the underlying normal distributions that define the profiles) (step 1), it is often of interest to explore associations between profile categories and covariates (e.g., age, sex) or outcomes (e.g., CASI-IRT score), the latter often called external variables in the latent profile literature. Because the fundamental output of LPA is only a probability of belonging to each profile for each individual, additional steps are required to do this association. The most common next step in the literature is to assign each individual to the profile with the highest posterior probability (also known as modal assignment) (step 2), followed by association analyses between the assigned class membership and external variables, such as a health outcome (step 3). However, this three-step approach can lead to bias and invalid confidence intervals because the second step treats the class membership as an observed, perfectly measured grouping variable, ignoring the potential uncertainty, i.e., classification error introduced in step 2 (Bolck et al., 2014; Vermunt, 2010). Thus, it is always recommended to adopt an analytic approach that accounts for the uncertainty in class membership assignments. It should be noted that estimating the association of a latent variable with external variables is still an area of active research. There are several papers that review and compare different approaches proposed over the last two decades to deal with external outcomes in LPA (Dziak et al., 2016, Collier et al., 2017) and latent class analysis (LCA) (Nylund-Gibson and Choi 2018, Nylund-Gibson 2019, Bakk and Kuha 2021). Note that LPA and LCA are similar, differing only in the nature of their indicator variables (LCA applies to categorical indicator variables), not in the way the latent variable is related to an external variable. Introducing and discussing each method is beyond the scope of this article. Current recommended practice is to use either the improved three-step maximum likelihood (ML)-based method (Vermunt, 2010) or Bolck, Croons, and Hagenaars (BCH) method (Asparouhov and Muthen, 2014; Bolck et al., 2004; Vermunt 2010), which adjust for the uncertainty in the latent profile assignment. These methods assume conditional independence of covariates and profile indicators given the latent variable. For recent advances and discussion of more complex models, such as assuming covariates have direct effects on observed profile indicators, multilevel or latent transition models, or more than one latent class variables, refer to Bakk et al., 2021, Nylund-Gibson et al., 2019, Bray and Dziak 2018, and Vermunt et al., 2021.

In our data analysis, LPA was performed in the Latent GOLD 6.0 software (Vermunt and Magidson 2021). The improved BCH method, which has been advocated as a preferred method for continuous external outcomes (Bakk et al., 2021), was used to relate latent profiles with the outcome CASI-IRT score adjusted for age, sex, race/ethnicity, BMI, education level, depressive symptoms, and self-rated health. Latent GOLD syntax for this analysis is provided in Supplemental material A2.3. We compared these results to those of the biased approach of ignoring the uncertainty in the class assignment. As a sensitivity analysis, we repeated this analysis with the ML-based instead of the BCH adjustment for the uncertainty in the latent-profile assignment using the same software. We also performed analyses of using each of the covariates as a predictor for latent profiles with the ML-based method, which can be done using the Latent GOLD "Step-3" module. P-values for all associations were reported based on the Wald-type test using robust standard error estimates (using robust standard error estimates is necessary as shown in Vermunt 2010 and Bakk et al., 2014 and is the default in Latent GOLD 6.0).

*3.3.1 LPA Illustration*

The distributions and correlations between four activity behaviors can be found in Supplemental Figure S2. In our analysis, the distributions of *ilr*-transformed variables were more skewed than those on the original scale (see Supplemental Figures S2 and S3). Thus, we dropped sleeping from the analysis to avoid the degenerate variance matrix and because our interest in the profiles is driven by the waking time activities (though sleeping is still implicitly included in the resulting profiles as the four proportions sum to one). We allowed the variance matrix to vary across profiles and fit a series of models with 2-6 latent profiles in Latent Gold (Vermunt and Magidson, 2021) with 160 randomly generated seed values and a maximum of 250 iterations for the E-M algorithm. No convergence issues were observed and the best solution based on log-likelihood value can be replicated in more than 10% of runs. LPA can be also performed in R with the package tidyLPA (Rosenberg et al., 2019) and Mplus (Muthén and Muthén (1998–2017)). A tutorial of using this R package and its comparison to Mplus can be found in Wardenaar 2021. Model fit statistics (Supplemental Table S3) were used as initial screening methods to select candidate models and then combined with each model's interpretability, minimal profile size to determine, and finally the likelihood ratio-based tests to determine the final number of profiles.

Supplemental Table S3 presents model fit statistics for models with 2-6 number of latent profiles. AIC, BIC, CAIC, ABIC, ICL-BIC favor the model with 6, 3, 3, 4, and 2 latent classes, respectively. The solution with 6 latent classes has the lowest AIC; however, Tein et al., 2013 noted that the AIC is not a reliable method for selecting the number of classes. The 6-class solution also has a minimal class size of only 46 observations (4.4% of the entire sample), which may indicate overfitting. Based on the bootstrap likelihood ratio test comparing the three-class and four-class solutions, the p-value of <0.001 indicates the four-class model fits the data better. Comparing the distributions of four activity behaviors across classes based on a four-class model (Figure 4) to that based on a three-class model (Supplemental Figure S4), we can see that the four-class model additionally identifies a group with lower than average sleeping time, as well as higher activity (standing and stepping) compared to profiles 3 and 4 with similar or lower sitting time, which could correspond to a clinically interesting 24HAC profile in this ACT cohort. Hence, the model with four latent profiles was chosen for further inferential analysis. The estimated means and variance-covariance matrices, and probability of observing each profile are presented in Supplemental Table S4. The profiles were labelled 1 to 4 according to their estimated mean sitting time per day (low to high) with each profile's unique characteristics summarized (mean (SD) time spent in each activity reported) as below.

Profile 1 ("Most active") accounted for 15.9% of the sample and was the most active group with the least sitting time (mean (SD) = 7.6 (1.4) hours/day), the highest standing time (5.8 (1.8) hours /day) and stepping time (1.9 (0.8) hours /day), and about average sleeping time (8.7 (1.1) hours /day).

Profile 2 ("Moderately active low sleepers") included 24.4% of the population, characterized by the lowest sleeping time (7.9 (1.0) hours /day). Interestingly, this group also had the second highest standing time (4.7 (1.1) hours /day) and stepping time (1.7 (0.5) hours /day), and second lowest sitting time (9.6 (1.1) hours /day).

Profile 3 ("Average activity") accounted for 40.3% of the population and represented time spent on all activity behaviors around the sample average (about 10.3 (1.3) hours /day sitting, 3.5 (0.9) hours /day standing, 1.3 (0.4) hours /day stepping, and 8.9 (0.6) hours /day sleeping).

Profile 4 ("Least active") included 19.4% of the sample and had the least standing and stepping time (about 2.4 (1.0) hours /day and 0.7 (0.3) hours /day, respectively), highest sitting time (about 12.1 (1.7) hours /day), and average sleeping time (about 8.8 (1.4) hours /day).

When each individual is assigned to the profile with highest posterior probability, Table 4 shows the descriptive statistics of sample characteristics across assigned memberships. Almost all factors listed are significantly associated with profile group assignment. Using the "Average activity" (profile 3) as the reference, more active groups (profile 1 and 2) are more likely to be female (especially for the "Most active" group, i.e., profile 1), relatively younger, have a lower BMI and better physical function, and experience slightly better sleep quality and cognitive function measured by CASI-IRT score. For the "Moderately active, low sleepers" (i.e., profile 2), besides the lowest median sleeping time, this group has the highest percentage of participants aged 65-74 and identifying as non-White. The "Least active" group, relative to the "Average activity" group, is older, more male, and has worse depressive symptoms, sleep quality, and physical and cognitive function. It should be noted that the descriptive means and standard errors in this table do not account for the uncertainty in the class membership assignments. The entropy statistic of the final solution was 0.55 and Supplemental Table S5 provides estimated probabilities of misclassification, both of which indicate a level of uncertainty involved in class assignments that should not be ignored in inferential analysis. After accounting for the classification errors, the last column of Table 4 offers p-values for bivariate associations of all considered characteristics with the latent profiles.

Table 5 provides the estimates for the effects of latent profiles on CASI-IRT scores with and without controlling for other covariates using the improved BCH method. Without any adjustment, the Wald test p-value for the association of latent profile with continuous CASI-IRT score is 0.033. With adjustment, the Wald test p-value becomes 0.88. The association estimates in Table 5 are also contrasted to that from linear regression analyses but with uncertainty in class membership assignments ignored. We see attenuated effects of the latent profiles on CASI-IRT scores and underestimated standard errors from the linear regressions, showing the fundamental problem of bias and invalid inference using the naïve approach of ignoring the uncertainty in the profile membership. For example, in the case of no adjustment, the linear regression estimates for the expected differences in CASI-IRT score between profile 3 ("average activity") and profile 4 ("least active") was 0.171 SD units (robust standard error 0.059 and p-value 0.005) with profile 3 having higher mean CASI-IRT score. In contrast, after accounting for misclassification, the difference became 0.239 SD unit (rSE 0.101 and p-value 0.017). Although the associations of the profiles with CASI-IRT scores from both analyses turned out to be nonsignificant with adjustment of other covariates, attenuated effects and underestimated standard errors in the approach that ignored the uncertainty in the latent profile assignment can be still observed. Results were very similar based on the ML-based analysis method to handle the uncertain profile assignment (data not shown).

## 4. Comparative Summary

We summarize the three analytical approaches to analyze 24HAC presented in Section 3 and highlight differences in the assumptions, research questions addressed, and other attributes of the associated regression approaches. An overall summary of ideas discussed is provided in Table 6A-B.

*Research questions and assumptions*

The main research question that both ISM and CoDA address regarding 24HAC data is the associated effect of time reallocation on the study outcome. Typically, ISM is focused on substituting time spent in one activity for another, whereas CoDA more naturally considers differences between two compositions. Both ISM and CoDA estimate the substitution effect in a standard regression framework and the usual assumptions would apply, e.g., for our ACT study example, usual linear model assumptions apply. ISM is generally conducted on the original time scale (e.g. hours of activity); whereas, CoDA models the proportion of time spent in each activity. Hence, CoDA is scale invariant, which implicitly means that the time information of 24HAC is irrelevant in CoDA, e.g., the amount of time spent on each behavior, which is an additional assumption that researchers should evaluate. In contrast to ISM and CoDA, LPA is a more exploratory method used to identify distinct latent subgroups with respect to activity profiles based on observed 24HAC data. Another aim of LPA is to explore the correlates of identified profiles and the associations of the profiles with outcomes. LPA assumes the observed data are sampled from a population composed of distinct subpopulations with heterogeneous distributions of activity behaviors of a day and models the data using a mixture of multivariate normal distributions. Hence, LPA results can be sensitive to the distributions of activity behaviors, while for ISM and CoDA, this assumption is not necessary if the outcome model is correctly specified. Secondary research questions of interest may relate to comparisons of descriptive statistics of the 24HAC by different population characteristics. We discuss this as a part of exploratory data analysis in the following section.

*Exploratory data analysis*

In any regression setting, exploratory data analysis is carried out to provide a summary of observed data and justification of the modeling assumptions needed for further analysis. By considering each activity behavior of the 24HAC univariately, we can plot the distribution of each behavior, e.g., in Supplemental Figure S2. Further bivariate association/stratification analysis can also be done to explore possible correlates of each activity behavior. Under CoDA, there are unique tools and descriptive statistics available. For example, ternary diagrams, e.g., Figure 1, can be used to visualize the distribution of 24HAC component behaviors. A compositional mean and variation matrix can be calculated overall and in subgroups as descriptive statistics to summarize central tendency and co-dependences of activity behaviors. By applying the *ilr*-transformation, statistical comparisons of compositional means across subgroups can also be done using standard multivariate analysis of variance, which we think is superior to the bivariate association analysis that considers each activity behavior univariately and does not account for the co-dependence in 24HAC data. For LPA, profile-specific estimated means and variance matrices for time spent in each activity from a final fitted model, as well as a boxplot plot (e.g., Figure 4), can describe the distributions of each activity behavior across profiles. Bivariate associations of covariates with the profiles can be explored by calculating the summary statistics of each covariate across the profiles, e.g., Table 4. However, it should be noted that directly performing bivariate association analysis with the assigned class and ignoring the class uncertainty can lead to biased and inaccurate results and should be verified using an adjusted regression method (Bolck et al., 2004; Vermunt, 2010). Some adjustment methods are recommended. See LPA Section 3.3. This uncertainty in the profile assignment makes exploratory data analysis more challenging in the context of LPA, since one may rely on the modal assignment, which is subject to misclassification.

*Multivariable regression analysis*

Both ISM and CoDA estimate the effects of time reallocations through standard multivariable regression models; however, they handle the collinearity between activity behaviors differently. ISM is formulated by including the total activity and all but one of the activity variables – the activity you will explore reallocating. In contrast, CoDA considers 24HAC as a composition and applies the isometric log-ratio (ilr) transformation to transform the compositional data into a set of continuous variables in a lower dimensional space that standard statistical approaches can work with. Although the procedure and the interpretation of the 24HAC model coefficients are more complicated, regarding 24HAC as a composition facilitates the comparison of health outcomes between any two compositions. Such comparisons need to be made within the main range of data to avoid extrapolation. Furthermore, although linear relationships are assumed between *ilr*-coordinates and an outcome, when results are anti-logged, the effects of time reallocations are nonlinear with respect to the amount of time reallocated, e.g., see Figure 2 and Figure 3 for nonlinear and asymmetric effects. ISM, in contrast, standardly estimates the linear effects of time reallocation; however, both ISM and CoDA have potential to be more flexible, e.g., by including higher order, spline terms, or extending to more complex semi-parametric or non-parametric models. Although not illustrated in detail in Section 3, standard model checking and diagnostic tools for multivariable regressions largely apply to both ISM and CoDA. For LPA, instead of estimating the effects of time reallocations, we can only estimate associations between an outcome and the identified profiles by using an approach that accounts for potential misclassification due to the uncertainty in the profile assignment, e.g., the improved three-step approach (Vermunt 2010) as in Table 5. Lastly, although our illustrations were with a continuous outcome variable, all the three methods are applicable when other common types of outcomes are used, e.g., dichotomous or time-to-event (Merkery et al., 2013, Mcgregor et al., 2020, Lythgoe et al., 2019). Again, it should be noted that it is more challenging to do model checking for LPA due to the uncertainty in the profile assignment. Each model diagnostic statistic would need to be adjusted for the potential for misclassification

*Statistical challenges*

If a zero value occurs for an individual for any behavior in the composition or it is of interest to predict an outcome based on a fitted CoDA regression model for a population with time spent in any behavior close to zero, CoDA will have numerical problems because of its reliance on log-transformations. Even for ISM and LPA, common zero values could lead to skewed distributions and could be influential on final results. If zero observations account for a very small proportion of the observed data for a given behavior, then one approach can be to consider these values as below the limit of detection. In such cases, it is common practice to replace the zero values with a small value relative to the observed data, e.g., ½ limit of detection, but this may not always be sensible. Alternatively, values could be imputed by statistical imputation tools (Martin et al., 2003, Palarea et al., 2007). If the occurrence of zero values is expected for one or more behaviors, CoDA may not be an appropriate approach. For example, if MVPA was considered as a component of the 24HAC for a given cohort with limited physical function, some individuals could have zero minutes/day in MVPA. In this case, CoDA could only be applied if the 24HAC is redefined as a composition of behaviors that all members typically engage in, such as by merging two or more activity behaviors into a single behavior (e.g., stepping, which would include most MVPA, along with lighter-intensity movement) before conducting CoDA.

Detecting and capturing non-linear effects of time reallocations could be a challenge to both ISM and CoDA. Developing models allowing for non-linear effects of the 24HAC exposure but with good interpretability is a direction of future research. LPA is a compelling method by which to summarize distinct patterns of activity behavior in a population, but this approach has a number of statistical challenges: including the uncertainty of the latent class assignment leading to non-standard procedures for statistical inference for the association of latent profiles with an outcome, reliance of current methods on multivariate normality, the need to drop one of the activity behaviors in the 24HAC due to the collinearity, and created profiles being unique to the specific cohort understudy.

**5. Conclusion**

The 24HAC is an important new paradigm by which to summarize activity behaviors. This approach naturally captures the multivariate nature of physical activity and sleep behaviors. The analytical approaches considered in this work each have specific advantages and limitations to consider. Which method works well for a given setting will depend on the primary scientific question of interest and structure in the data. ISM is a simple and easy to interpret model to apply when linear associations are of interest and the central question relates to the association of substituting one activity for another with changes in outcome. CoDA allows for more direct assessment of the associated difference in outcome between different compositions of activity behaviors in the 24HAC and inherently models a non-linear association with changes in any one component of activity. LPA shifts the focus to detecting different sub-populations whose patterns of 24HAC differ the associated differences in outcome between these subpopulations. The motivating research question would largely determine which of these methods would be best suited for the analysis. It may also be useful to apply more than one method to a setting.

In our ACT Study illustration, we saw little to no cross-sectional association between 24HAC and cognition measured by the CASI-IRT score. The ACT-AM sub-study cohort represented a somewhat healthier subset of the larger ACT Study cohort (Rosenberg et al., 2020), as well as overall in that it excludes those with an existing dementia diagnosis, which may have led to less heterogeneity and lower power to detect associations. Analyses were also limited by being a complete case analysis of those with four or more valid days of wear, which represented a 91% subset of those who wore the ActivPal device (Rosenberg et al., 2020). Additionally, the analyses were potentially limited by being cross-sectional. Follow-up on ACT participants is ongoing, which will inform future research on longitudinal 24HAC patterns and associations with cognition and other health outcomes. A further limitation is that we used a measure of time in bed as a proxy for sleep. Time in bed could include wakeful sedentary time, such as reading in bed, which could obscure the specific associations between sleep and health outcomes. Future work in the ACT study is also underway to objectively measure sleep as part of the 24HAC in order to examine whether 24HAC compositions that delineate sleep may lead to differences in cognitive outcomes longitudinally.

Under a close-to-null association, the association in the CoDA model is approximately linear and the ISM and CoDA models provided very similar results in the activity substitution analyses. These models would differ more under stronger associations. The LPA provided a way to consider outcome differences between groups with different patterns of activity. In the unadjusted analyses, those that were the most active were seen to have significantly higher CASI-IRT scores; however, the association was no longer significant in the multivariable model. LPA regression results were subject to attenuation bias and inappropriately narrow confidence intervals when the results were not adjusted for the uncertainty in the profile assignment. This lack of adjustment is common in the published literature; however, statistical software is available to make this adjustment straightforward. LPA regression models are attractive because of their interpretability; however, they can also be subject to labeling bias. That is, the labels given to the different profiles can be misleading in that they could be suggestive of larger differences than exist in the data and also may inadequately capture all the ways in which the profiles differ, with respect to the distribution of the indicator variables. Care should be given that the labels are not over simplifying or misleading relative to the observed between-profile differences across the indicator variables.

The regression models that included components of the 24HAC directly, namely ISM and CoDA, provide limited flexibility in the modeled association between the 24HAC and the outcome. Future work is needed to develop more flexible regression models to study the 24HAC and its relationship with outcomes of interest. More work is also needed to improve current statistical approaches for LPA. Vermunt 2010 found that for latent class analysis (LCA), in the case of small samples and low separation between classes, ignoring the uncertainty from the estimation of a latent class model can also lead to biased standard errors of associations with external variables even if the uncertainty in class assignments has been accounted for. Bakk et al., 2014 studied this issue in more depth. The finding likely holds in LPA as well. In this paper, we only covered methods dealing with the uncertainty in LPA class assignments. How to account for both layers of uncertainty could be a direction of future research. Other approaches not considered here include functional principal components analysis of activity profiles measured by accelerometers (Xu et al., 2019, Xiao et al., 2022) and latent class models of longitudinal biomarkers (Proust-Lima et al., 2014, 2022). A recent review of CoDA, also discusses other analytical approaches for compositional data, including advocating for alternative, simpler transformations to the *ilr*-transformation (Greenacre et al., 2022).

The 24HAC is an exciting new paradigm to study how differences in activity behavior affect cognitive and other clinical outcomes. The ISM, CoDA, and LPA methods provide three useful approaches that each, with appropriate application and interpretation, can lead to useful insights regarding the association of 24HAC with outcomes. Future work in methodology to expand the flexibility in available models of 24HAC is necessary in order to enhance the ability to understand the potentially complex nature of these associations.

**Acknowledgements:** This research was funded by the National Institute on Aging (U19AG066567). Data collection for this work was additionally supported, in part, by prior funding from the National Institute on Aging (U01AG006781). All statements in this report, including its findings and conclusions, are solely those of the authors and do not necessarily represent the views of the National Institute on Aging or the National Institutes of Health. We thank the participants of the Adult Changes in Thought (ACT) study for the data they have provided and the many ACT investigators and staff who steward that data. You can learn more about ACT at: https://actagingstudy.org/

**Table 1 Descriptive statistics for the ACT 24-hour Activity Study (N=1034).[1]**

| | Group | Overall |
|---|---|---|
| **Categorical variables, n (%)** | | |
| Age category, years | 65-74 | 433 (41.9) |
| | 75-84 | 425 (41.1) |
| | 85+ | 176 (17.0) |
| Sex | Male | 457 (44.2) |
| | Female | 577 (55.8) |
| Race[2] | White | 933 (90.2) |
| | Black | 18 (1.7) |
| | Asian | 30 (2.9) |
| | Native Hawaiian/Pacific Islanders | 2 (0.2) |
| | Other race | 49 (4.7) |
| Hispanic ethnicity[2] | Yes | 14 (1.4) |
| | No | 1017 (98.4) |
| Self-rated health[3] | Excellent/Very good/Good | 952 (92.1) |
| | Fair/Poor | 82 (7.9) |
| Physical function score | <= 10 | 723 (69.9) |
| | >10 | 243 (23.5) |
| | Missing | 68 (6.6) |
| Ability to walk half a mile[2] | No difficulty | 771 (74.6) |
| | Some difficulty | 149 (14.4) |
| | A lot difficulty or Unable | 110 (10.6) |
| Time in bed, hours[4] | <6 | 17 (1.6) |
| | 6-9 | 669 (64.7) |
| | 9+ | 348 (33.7) |
| Sleep quality | Fair, poor, or very poor | 342 (33.1) |
| | Good or very good | 596 (57.6) |
| | Missing | 96 (9.3) |
| *Continuous variables, Mean (SD)* | | |
| Age, years | - | 77.2 (7.0) |
| Body mass index[2], kg/m$^2$ | - | 27.1 (4.9) |
| Years of education | - | 16.8 (2.8) |
| Depressive symptoms (CES-D Score)[2] | - | 3.6 (3.9) |
| Cognition (CASI-IRT score) | - | 0.61 (0.69) |
| Sit time[5], hours/day | - | 10.0 (2.0) |
| Stand time[5], hours/day | - | 4.0 (1.6) |
| Step time[5], hours/day | - | 1.4 (0.7) |
| Sleep time[5], hours/day | - | 8.5 (1.1) |
| Total time[5], mins/day (median [IQR[1]]) | - | 1440 [1436, 1445] |

[1] ACT, Adult Changes in Thought; IQR, interquartile range. min, minute
[2] Missing data: ability to walk ½ mile n=4; body mass index n = 20; depressive symptoms n = 9, Hispanic ethnicity n=3; race n=2. Percents may not add up to 100%.
[3] Self-rated health ranges from 1=excellent, 5=poor
[4] Based on sleep log, the arithmetic mean of sleep durations over valid wear days, calculated as difference from current day's in-bed time to next day's out-bed time.
[5] Each behavior is treated here as a univariate variable and time spent on each behavior per day equals the arithmetic mean of the time spent on that behavior over valid wear days.

**Table 2. Isotemporal Substitution Models (ISM) in the overall sample and by mean step time subgroups (> or ≤ 60 mins/day). Change in the CASI-IRT score for a 30-minute/day time reallocation from the row label behavior to the behavior indicated by the column label (N = 1000).[1]**

|  |  | Sit | Stand | Step | Sleep |
|---|---|---|---|---|---|
|  | Activity Replaced | β [95% C.I.] | β [95% C.I.] | β [95% C.I.] | β [95% C.I.] |
| Overall (N=1000) | Sit | - | -0.006 [-0.020, 0.007] | 0.027 [-0.011, 0.064] | 0.000 [-0.019, 0.019] |
|  | Stand | 0.006 [-0.007, 0.020] | - | 0.033 [-0.011, 0.078] | 0.007 [-0.014, 0.028] |
|  | Step | -0.027 [-0.064, 0.011] | -0.033 [-0.078, 0.011] | - | -0.027 [-0.067, 0.014] |
|  | Sleep | -0.000 [-0.019, 0.019] | -0.007 [-0.028, 0.014] | 0.027 [-0.014, 0.067] | - |
| Step time > 60 min/day (N=270) | Sit | - | -0.009 [-0.026, 0.008] | 0.013 [-0.033, 0.058] | -0.001 [-0.024, 0.023] |
|  | Stand | 0.009 [-0.008, 0.026] | - | 0.021 [-0.031, 0.073] | 0.008 [-0.018, 0.035] |
|  | Step | -0.013 [-0.058, 0.033] | -0.021 [-0.073, 0.031] | - | -0.013 [-0.063, 0.036] |
|  | Sleep | 0.001 [-0.023, 0.024] | -0.008 [-0.035, 0.018] | 0.013 [-0.036, 0.063] | - |
| Step time ≤ 60 min/day (N=730) | Sit | - | -0.002 [-0.028, 0.024] | 0.037 [-0.172, 0.247] | 0.001 [-0.031, 0.033] |
|  | Stand | 0.002 [-0.024, 0.028] | - | 0.039 [-0.181, 0.260] | 0.003 [-0.034, 0.041] |
|  | Step | -0.037 [-0.247, 0.172] | -0.039 [-0.260, 0.181] | - | -0.036 [-0.246, 0.174] |
|  | Sleep | -0.001 [-0.033, 0.031] | -0.003 [-0.041, 0.034] | 0.036 [-0.174, 0.246] | - |

[1]All models were adjusted age group (65-74, 75-84, 85+), non-Hispanic White (Yes, No), body mass index, depressive symptom CES-D score, and self-rated health condition. 34 observations were excluded from the entire sample due to missing data in the covariates. Confidence intervals crossing zero are not statistically significant.

**Table 3. CoDA pivot coordinates ($z_1$) parameter estimates for the multivariable regression of CASI-IRT on 24HAC profiles for sleep, sit, stand, and step (N=1000).[1]**

|  | Estimate (95% C.I.) | P-value |
|---|---|---|
| Sit vs Remaining | -0.00 [-0.17, 0.16] | 0.954 |
| Stand vs Remaining | -0.05 [-0.18, 0.07] | 0.384 |
| Step vs Remaining | 0.08 [-0.02, 0.19] | 0.114 |
| Sleep vs Remaining | -0.03 [-0.24, 0.19] | 0.814 |

[1]Estimates in each row derived from fitting a linear regression with CASI-IRT as the outcome, and *ilr*-coordinates (each behavior in turn being the numerator in the pivot coordinate) as predictors, adjusted for age groups (65-74, 75-84, 85+), sex, non-Hispanic White (Yes, No), body mass index, depressive scores, and self-rated health conditions. 34 participants were not included due to missing values on covariates. Confidence intervals crossing zero are not statistically significant.

**Table 4. Participant characteristics across latent profiles according to modal assignment, i.e., every individual is assigned to the class with highest posterior probability (N=1034).**

| | Profiles | | | | |
|---|---|---|---|---|---|
| | Most Active | Moderately Active Low sleeper | Average activity | Least Active | |
| N | 132 | 257 | 453 | 192 | |
| **Sit** (hours/day) (median [IQR]) | 7.0 [6.4, 7.8] | 9.5 [8.5, 10.7] | 10.3 [9.4, 11.1] | 12.6 [11.4, 13.3] | |
| **Stand** (hours/day) (median [IQR]) | 6.3 [5.4, 7.3] | 4.9 [4.2, 5.5] | 3.5 [2.9, 4.1] | 2.2 [1.5, 2.7] | |
| **Step** (hours/day) (median [IQR]) | 2.1 [1.2, 2.7] | 1.8 [1.4, 2.1] | 1.3 [1.0, 1.6] | 0.7 [0.5, 0.8] | |
| **Sleep** (hours/day) (median [IQR]) | 8.5 [7.9, 9.4] | 7.8 [7.3, 8.4] | 8.9 [8.5, 9.3] | 8.7 [7.8, 10.0] | |
| **Categorical variables, n (%)** | | | | | p-value[1] |
| **Female** | 89 (67.4) | 140 (54.5) | 253 (55.8) | 95 (49.5) | 0.025 |
| **Age** | | | | | <0.001 |
|   65-74 | 64 (48.5) | 132 (51.4) | 185 (40.8) | 52 (27.1) | |
|   75-84 | 54 (40.9) | 97 (37.7) | 196 (43.3) | 78 (40.6) | |
|   85+ | 14 (10.6) | 28 (10.9) | 72 (15.9) | 62 (32.3) | |
| **Hispanic ethnicity or Asian /Black/other race minority race[2]** | 13 (9.8) | 35 (13.7) | 41 (9.1) | 20 (10.5) | 0.30 |
| **Self-rated health[3]** = Excellent/Very good/good | 119 (90.2) | 245 (95.3) | 431 (95.1) | 157 (81.8) | <0.001 |
| **Physical function[4,5]** = >10 | 39 (32.0) | 73 (30.3) | 103 (23.6) | 28 (16.9) | 0.007 |
| **Ability to walk half a mile[6]** | | | | | <0.001 |
|   No difficulty | 108 (81.8) | 223 (86.8) | 357 (79.0) | 83 (43.9) | |
|   Some difficulty | 17 (12.9) | 22 (8.6) | 61 (13.5) | 49 (25.9) | |
|   A lot difficulty or unable | 7 (5.3) | 12 (4.7) | 34 (7.5) | 57 (30.2) | |
| **Time in bed[3]** | | | | | <0.001 |
|   <6 hrs | 4 (3.0) | 8 (3.1) | 0 (0.0) | 5 (2.6) | |
|   6-9 hrs | 88 (66.7) | 224 (87.2) | 252 (55.6) | 105 (54.7) | |
|   9+ hrs | 40 (30.3) | 25 (9.7) | 201 (44.4) | 82 (42.7) | |
| **Sleep quality[4]** = good, very good | 83 (69.2) | 158 (65.8) | 265 (64.8) | 90 (53.3) | 0.02 |
| **CASI-IRT category** | | | | | 0.10 |
|   <= 0 | 23 (17.4) | 40 (15.6) | 84 (18.5) | 51 (26.6) | |
|   (0, 1] | 68 (51.5) | 132 (51.4) | 215 (47.5) | 91 (47.4) | |
|   >1 | 41 (31.1) | 85 (33.1) | 154 (34.0) | 50 (26.0) | |
| **Continuous variable, mean (SD)** | | | | | |
| **Body mass index[4]**, kg/m$^2$ | 25.4 (4.21) | 26.0 (4.3) | 27.1 (4.8) | 29.5 (5.5) | <0.001 |
| **Years of education** | 16.7 (2.9) | 17.3 (2.6) | 16.8 (2.7) | 16.2 (3.1) | 0.002 |
| **Depressive symptoms (CES-D Score)[4]** | 3.6 (4.3) | 3.3 (3.4) | 3.3 (3.6) | 4.7 (4.7) | <0.001 |
| **CASI-IRT Score** | 0.63 (0.67) | 0.66 (0.64) | 0.63 (0.70) | 0.46 (0.72) | 0.013 |

[1] Wald-type test based on robust standard errors; p-values are for testing the univariate association of each covariate with latent profiles by accounting for potential misclassification in class membership assignment (Vermunt 2010).
[2] Self-rated health ranges from 1=excellent, 5=poor
[3] Based on sleep log. For each subject, this variable equals the arithmetic mean of sleep durations over valid wear days. The sleep duration was the difference from current day's in-bed time to next day's out-bed time.
[4] Missing data: race/ethnicity N = 5; physical function N = 68; ability to walk half a mile N = 4; sleep quality N = 96; body mass index N = 20; depressive symptoms N = 9.

**Table 5. Association of latent profiles with the outcome CASI-IRT score: without adjustment for the potential profile misclassification due to the uncertainty in latent profile assignment versus an approach that adjusted for the potential misclassification (improved BCH method).[1]**

|  | No adjustment for misclassification | | | Adjustment for misclassification | | |
|---|---|---|---|---|---|---|
| **Without any adjustment (N=1034)** | β | Robust SE[3] | p-value[2] | β | Robust SE[4] | p-value[2] |
| Profile 3 "average activity group" | (ref) | (ref) | (ref) | (ref) | (ref) | (ref) |
| Profile 1 "most active group" | 0.002 | 0.067 | 0.98 | -0.048 | 0.096 | 0.62 |
| Profile 2 "moderately active lower sleeper" | 0.032 | 0.052 | 0.53 | 0.018 | 0.099 | 0.86 |
| Profile 4 "least active group" | -0.171 | 0.059 | 0.005 | -0.239 | 0.101 | 0.017 |
| **With adjustment of covariates[1] (N=1000)** | β | Robust SE[3] | p-value[2] | β | Robust SE[4] | p-value[2] |
| Profile 3 "average activity group" | (ref) | (ref) | (ref) | (ref) | (ref) | (ref) |
| Profile 1 "most active group" | -0.048 | 0.063 | 0.43 | -0.069 | 0.090 | 0.44 |
| Profile 2 "moderately active, lower sleeper" | -0.020 | 0.050 | 0.69 | -0.028 | 0.098 | 0.78 |
| Profile 4 "least active group" | -0.011 | 0.057 | 0.85 | -0.027 | 0.104 | 0.80 |

[1]Covariates adjusted: age groups (65-74, 75-84, 85+), sex, non-Hispanic White (Yes, No), body mass index, depressive scores, and self-rated health conditions
[2]P-values are based on Wald test using robust standard error estimates.
[3]Robust standard error (SE) is used to allow for potential heteroscedasticity in residual variance. It is usually larger than the usual standard error that assumes constant residual variance. However, in our case, the difference between the two is very small and hence only robust standard errors are reported.
[4]Using robust standard error is necessary for valid statistical inference as shown in Vermunt 2010 and Bakk et al., 2014.

**Table 6A**. Comparative summary of three methods to analyze 24HAC: Isotemporal substitution (ISM), Compositional Data Analysis (CoDA), and Latent Profile Analysis (LPA).

|  | ISM | CoDA | LPA |
|---|---|---|---|
| **Research questions of interest** | Effect of activity substitution | Effect of changes in 24HAC profiles, effect of activity substitution or more general | Exploratory summary of distinct activity subgroups (latent class identification) |
| **Assumptions** | Linear model, with possibility for non-linear terms for individual activities | Linear on transformed scale; scale invariance (i,e., relative proportion, not absolute quantity is relevant scale) | Within each latent class, activity behaviors (or their transformation) are assumed to be multivariate normal |
| **Exploratory data analysis** | | | |
| Data visualization | Individual activity summaries of descriptive statistics; e.g., box plots for each activity | Compositional plots with each behavior on a separate axis; e.g., Ternary plots | Plots of activity by profile groups; e.g. Grouped violin or boxplots by profile |
| Bivariate associations/stratification analysis | Analyses done separately for each activity | Compute mean composition in each subgroup; compositional two-sample t-test/ANOVA | Compute descriptive statistics of baseline factors by latent profiles; testing needs to account for uncertainties (requires specialized software) |
| **Multivariable regression analysis** | | | |
| Implementation | Standard regression, easy to apply | Isometric log-ratio transformation of 24HAC, standard regression | Specialized regression method that account for class uncertainty |
| Applicable for standard regression outcome models | Yes | Yes | Yes |
| Typical comparisons of different activity profiles | Substitution of one activity for another | Increasing one activity while proportionately decreasing the others, and substitution | Comparison of the latent profiles only |

**Table 6B. Statistical features and challenges of three methods to analyze 24HAC: Isotemporal substitution (ISM), Compositional Data Analysis (CoDA), and Latent Profile Analysis (LPA).**

|  | ISM | CoDA | LPA |
|---|---|---|---|
| **Interpretability** | Easy to interpret; harder when extended to nonlinear ISM | Not easy; requires understanding of how CoDA works | Not difficult; but be aware of uncertainty in class membership assignments |
| **Multicollinearity** | No perfect collinearity, but analysis could still suffer from high correlations between some activities | No perfect collinearity after isometric log-ratio transformation, but analysis could still suffer from high correlations among transformed data | Need to consider dropping one activity or apply isometric log-ratio transformation before running LPA |
| **Zero activity values** | Not an issue when model is correctly specified | Lead to numerical issues | Lead to skewed profile indicators and potential violation of LPA assumptions or need for a LPA method that allows for non-Gaussian indicators |
| **Non-linear association** | Generalizations available | Already nonlinear, could be more flexible | Not applicable |

**Figure 1. Ternary diagrams for each of the possible sub-compositions of 3 activities, with the CASI-IRT score represented with a color gradient scale (negative scores in blue, positive scores in orange) (N=1034).** For each ternary diagram, the vertices represent three behaviors of the composition, and each side of the triangle is an axis representing values for each behavior, ranging from 0% to 100%. Points that lie close to a vertex have high percentages of the behavior represented by that vertex, whereas points lying in the center of the triangle have equal percentages of all three behaviors. For example, for the solid black point on the diagram (A), we can read the percentage of each behavior for that point by drawing a line to each axis as illustrated by the black dashed lines (sit = 49%, stand = 38%, and step = 13%).

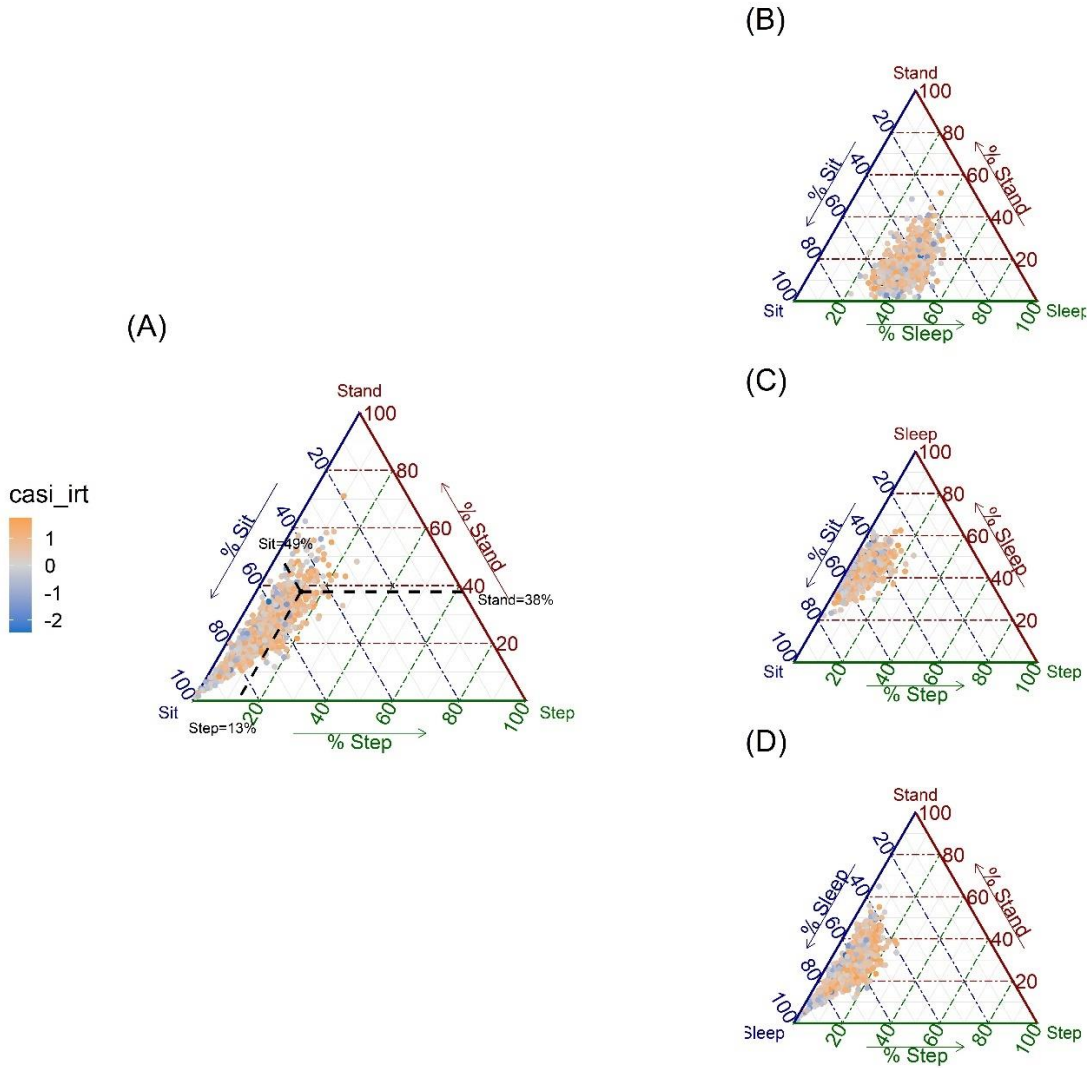

**Figure 2. Predicted difference in outcome from CoDA when increasing one of sleep, sit, stand step, while proportionally decreasing each of the other 3 behaviors. N = 1000[1].** Each one of the four subgraphs presents a curve of the predicted difference in mean outcome associated with increasing/decreasing time/day on the behavior (indicated in the header of the subgraph) by Δ mins, while accounting for that difference by proportionally decreasing/increasing time/day spent on other behaviors by a common factor. The curve presented in each graph is estimated based on the fitted model with CASI-IRT score as the outcome, and with *ilr*-coordinates as predictors in which the behavior (indicated in the header of the subgraph) as the numerator behavior in the pivot coordinate, adjusted for age groups (65-74, 75-84, 85+), sex, non-Hispanic White (Yes, No), body mass index, depressive symptoms score, and self-rated health condition. The shaded area in each subgraph corresponds to pointwise 95% confidence intervals, where overlap with the horizontal line at 0 indicates a null association. The baseline composition for each of these analyses equals to the compositional mean in the sample, i.e., 10.2h (42.6%) sit, 3.7h (15.3%) stand, 1.2h (5.2%) step and 8.8h (36.9%) sleep. 34 subjects were not included due to missing values on covariates.

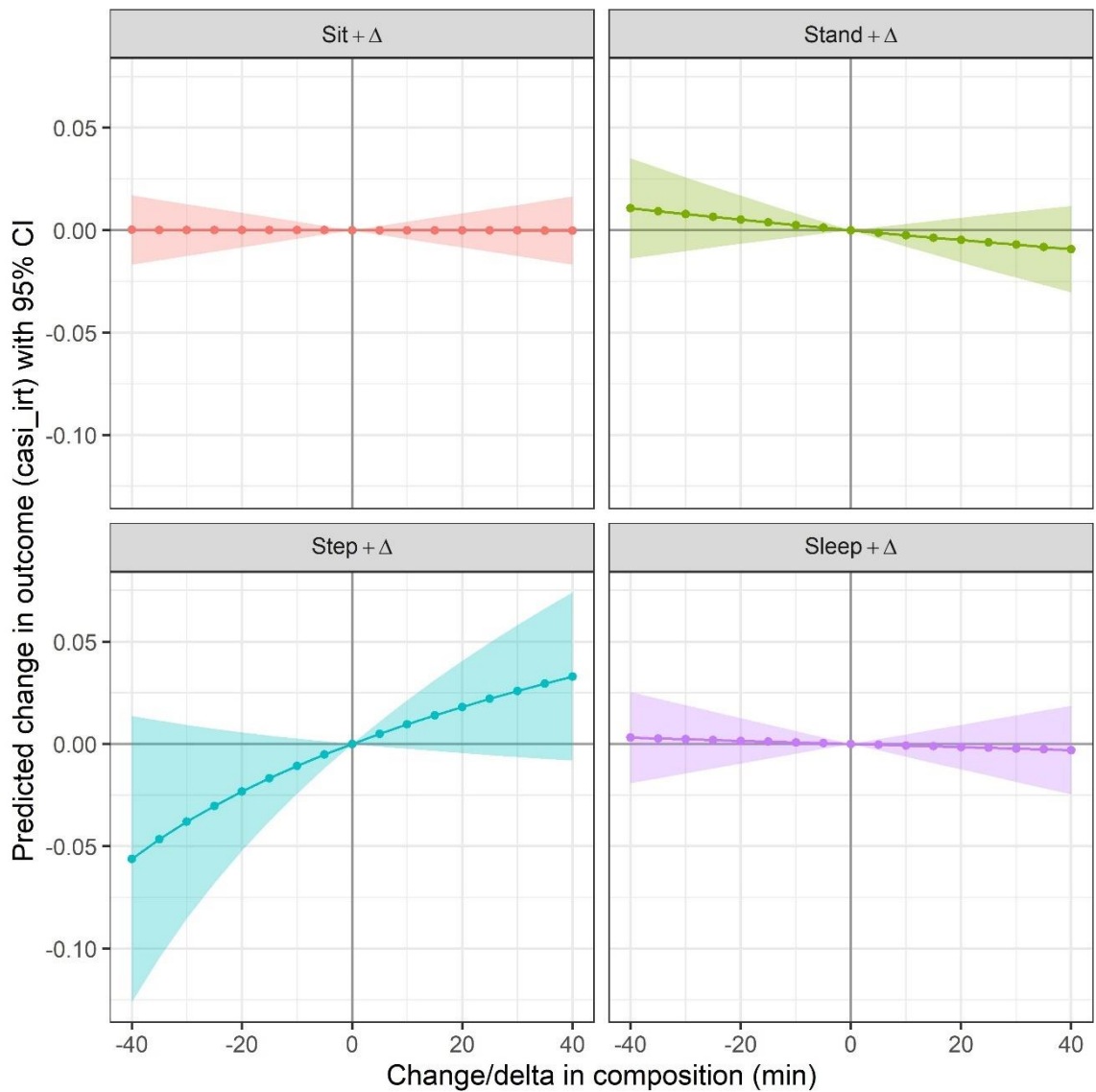

**Figure 3. Predicted difference in outcome from CoDA when reallocating time from one behavior to another, for all possible pairwise reallocations. N=1000**.[1] Each one of 16 subgraphs presents a curve of the predicted difference in mean outcome for increasing time/day on one behavior (indicated in the column header) by Δ mins, while decreasing time/day on another behavior (indicated in the row header i.e., grey vertical bar on the right) by the same amount of time. The curve presented in each graph is estimated based on the fitted model with CASI-IRT score as the outcome, and with any *ilr*-coordinates as predictors in which the behavior (indicated in the header of the subgraph) as the numerator behavior in the pivot coordinate, adjusted for age group (65-74, 75-84, 85+), sex, non-Hispanic White (Yes, No), body mass index, depressive symptoms score, and self-rated health condition. The shaded area in each subgraph corresponds to pointwise 95% confidence intervals. The baseline composition equals to the compositional mean in the sample, i.e., 10.2h (42.6%) sit, 3.7h (15.3%) stand, 1.2h (5.2%) step and 8.8h (36.9%) sleep. 34 participants were not included due to missing values on covariates.

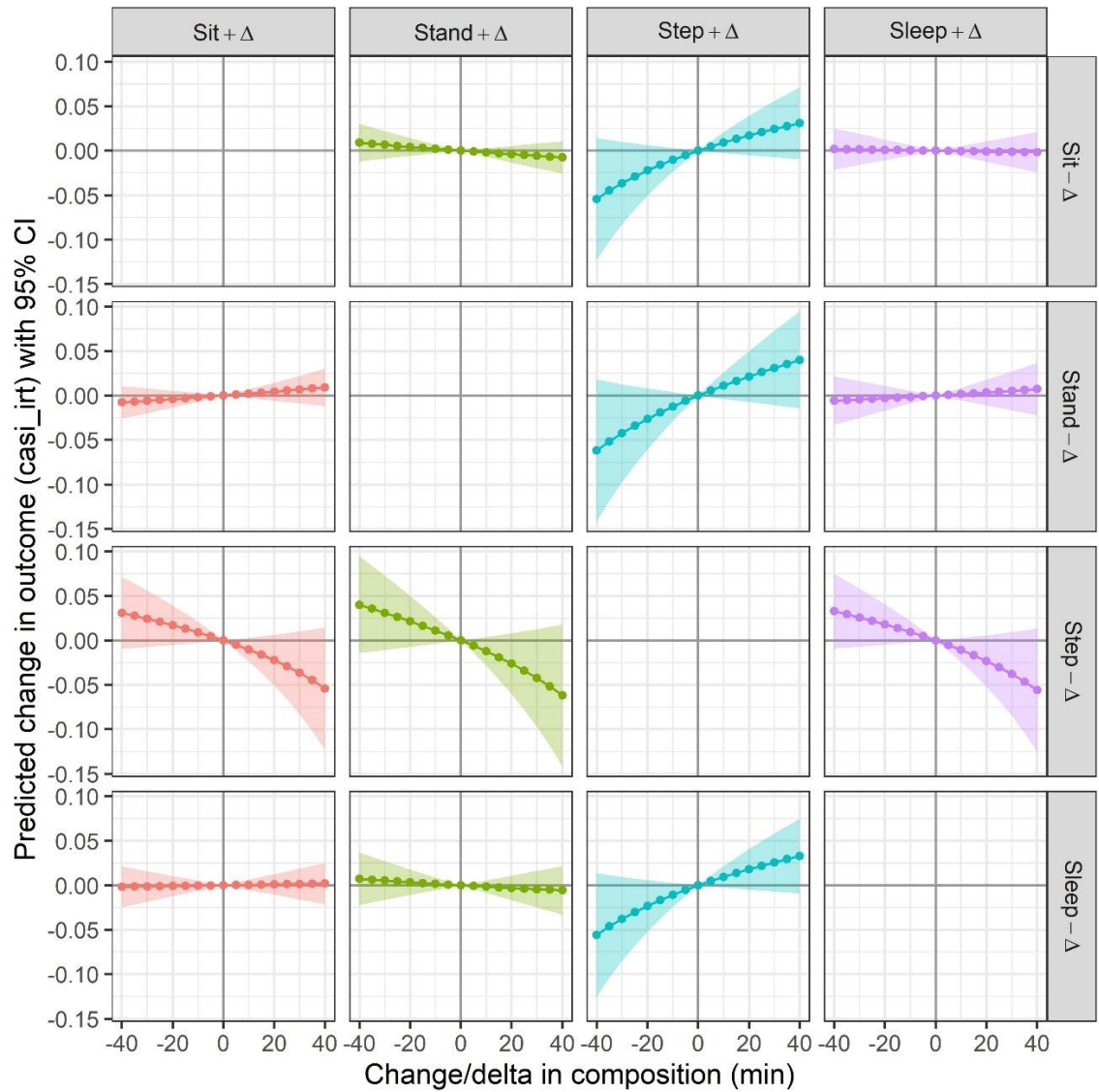

**Figure 4. Fitted 24HAC profiles from latent profile analysis (4-class solution), where each individual is assigned the class with maximum probability. The boxplots presents sample quartiles (N=1034).** The number (%) of subjects in profile 1-4 is, respectively, 132 (12.8%), 257 (24.9%), 453 (43.8%) and 192 (18.6%).

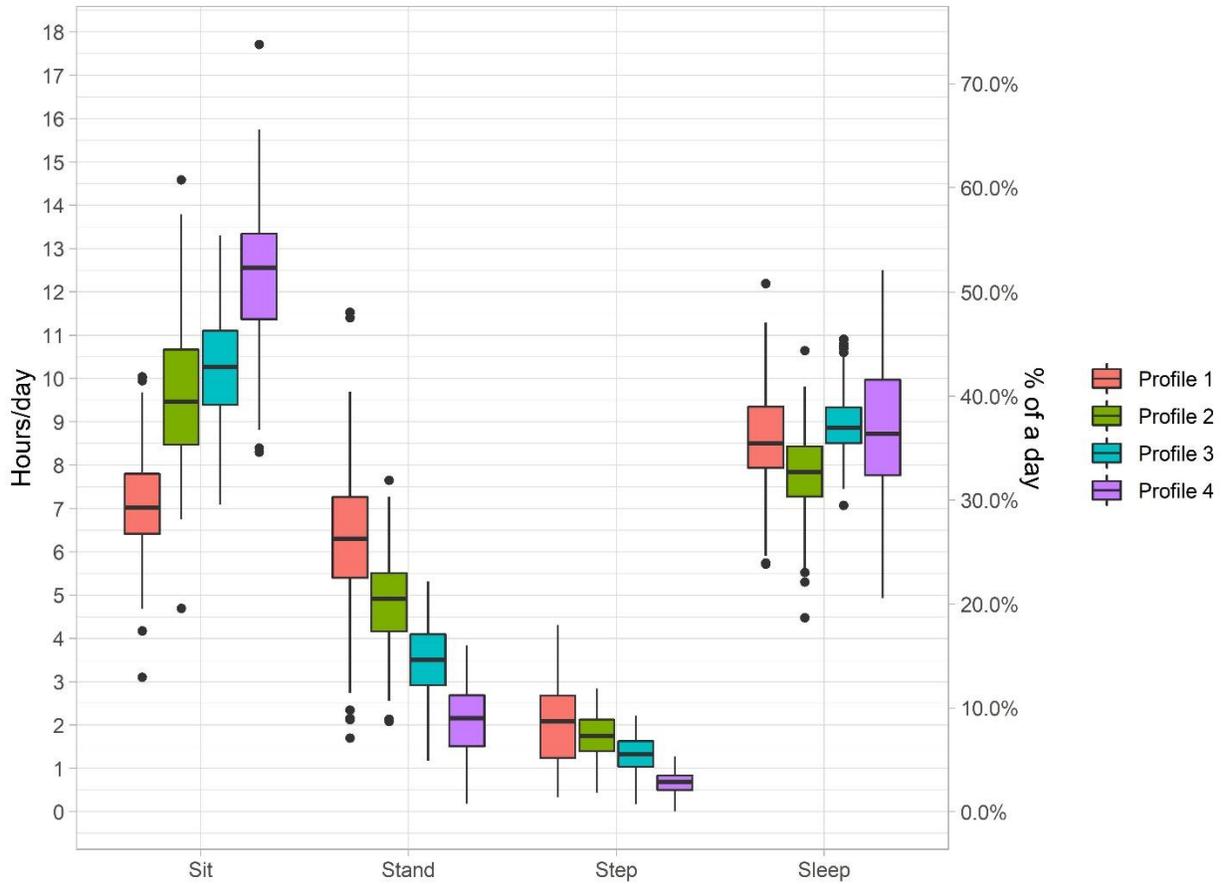

**Supplemental Web Materials for**

**Analysis of the 24-Hour Activity Cycle: An illustration examining the association with cognitive function-in the Adult Changes in Thought (ACT) Study**

Yinxiang Wu[1,2], Dori E. Rosenberg[1], Mikael Anne Greenwood-Hickman,[1] Susan M. McCurry[2], Cecile Proust-Lima[3], Jennifer C. Nelson[1], Paul K. Crane[2], Andrea LaCroix,[4] Eric B. Larson,[2] Pamela A. Shaw[1]*

1. Kaiser Permanente Washington Research Institute, Seattle, WA, USA
2. University of Washington, Seattle, WA, USA
3. Bordeaux Population Health Research Center, INSERM, Bordeaux, France
4. Herbert Wertheim School of Public Health and Human Longevity Science, University of California, San Diego, CA, USA

## A1. Compositional Data Analysis (CoDA) Supplemental Methods and Results

### A1.1 Definition of "addition" and "multiplication" operations in the simplex space and their connection to the definition of compositional mean

The collection of all D-part compositions $S^D = \{[x_1,..,x_D]: x_i > 0 \ (i = 1, ..., D), x_1 + \cdots + x_D = 1\}$ is called the simplex space. Aitchison (1986) defined two fundamental operations, namely perturbation and power operations, in the simplex space $S^D$ analogous to the addition and multiplication operations in real space. The perturbation operator denoted as $\oplus$ can be used to characterize 'difference' between compositions or change from one composition to another. In formula, the perturbation operation is defined as follows.

Consider two compositions $\mathbf{x} = [x_1, ..., x_D]$ and $\mathbf{y} = [y_1, ..., y_D]$,
$$x \oplus y = C[x_1 y_1, ..., x_D y_D] = [x_1 y_1, ..., x_D y_D]/(x_1 y_1 + ... + x_D y_D)$$
where $C$ is the so-called *closure* operation which divides each component of a composition by the sum of its components so as to scale the composition to the constant sum 1. It is not hard to see that the identity perturbation induced by this operation is $[\frac{1}{D}, ..., \frac{1}{D}]$ (meaning that every composition perturbed by this identity will remain the same, playing the same role as 0 in the real space) and the inverse of a perturbation denoted as $\ominus$ is given by $x \ominus y = C[\frac{x_1}{y_1}, ..., \frac{x_D}{y_D}]$. As as consequence, the change from $\mathbf{x}$ to $\mathbf{y}$ is expressed by the perturbation $\mathbf{y} \ominus \mathbf{x}$.

The operation analogous to scalar multiplication in real space is the power operation denoted as $\otimes$. For any real number a in R, and any composition $\mathbf{x}$ in $S^D$, the $a$-power transform of $\mathbf{x}$ is defined to be
$$a \otimes x = C[x_1^a, ..., x_D^a]$$

By drawing analogy with the definition of sample mean in real space i.e. mean $= \frac{1}{n}\sum(x_1 + \cdots + x_n)$, we can define the compositional mean in a similar way but by using the operations defined on the simplex space $S^D$. Suppose we observe n compositions in $S^D$, labelled as $x_1, ..., x_n$. By applying the same sample mean formula with replacement of every addition and multiplication operation in real space by the perturbation and power operation, we get $(\frac{1}{n}) \otimes (x_1 \oplus x_2 \oplus ... \oplus x_n)$ which is exactly the definition of compositional mean provided in the main text. Furthermore, if the "Aitchison distance" (Aitchison et al., 2000) is used as a metric of distance on the simplex space, it can be further shown that the compositional mean minimizes the "Aitchison distance" between all data points.

### A1.2 Sequential binary partition procedure to create ilr-coordinates and examples

The sequential binary partition process (SBP) (Egozcue et al., 2005) makes it easy to create *ilr*-coordinates. In the first step of the process, all parts are split into two groups and a coordinate is defined to be the log-ratio between the geometric mean of the two groups. In the following steps, each group in the previous ratio, is in turn split into two groups and the log-ratio between the geometric mean of the two groups is another coordinate, and the process continues until the two groups in the ratio each have a single component. A typical *ilr*-transformation for 24HAC keeps a single behavior in the numerator at each step, as shown in the example (equation 2) in main text. This process can be visualized as the table below.

| Level of partition | Sit | Stand | Step | Sleep | r | s |
|---|---|---|---|---|---|---|
| 1 | + | - | - | - | 1 | 3 |
| 2 |  | + | - | - | 1 | 2 |
| 3 |  |  | + | - | 1 | 1 |

r: the size of numerator group (labelled as "+"); s: the size of denominator group (labelled as "-")

The formula to calculate the *ilr*-coordinate resulted from each level of partition is $z = \sqrt{\frac{rs}{r+s}} \ln\left(\frac{N}{D}\right)$ where $N, D$ represent, respectively, the geometric mean of the activities involved in the numerator and denominator group.

Another other example of *ilr*-coordinates is as follows:

$$z_1 = \ln \frac{(Stand \times Step)^{1/2}}{(Sit \times Sleep)^{1/2}}$$

$$z_2 = \sqrt{\frac{1}{2}} \ln \frac{Stand}{Step}$$

$$z_3 = \sqrt{\frac{1}{2}} \ln \frac{Sit}{Sleep}$$

The table for the sequential binary process of generating this set of *ilr*-coordinates is

| Level of partition | Sit | Stand | Step | Sleep | r* | s* |
|---|---|---|---|---|---|---|
| 1 | - | + | + | - | 2 | 2 |
| 2 |  | + | - |  | 1 | 1 |
| 3 | + |  |  | - | 1 | 1 |

* r: the size of numerator group (labelled as "+"); s: the size of denominator group (labelled as "1")

The specific set of *ilr*-coordinates can be chosen to capture a particular comparison of geometric means to aid interpretation of that coordinate. For example, if the goal is to estimate the effect of increasing time in both stand and step while simultaneously decreasing time in sit and sleep on an outcome, then by regressing the outcome on the $z_1$, $z_2$ and $z_3$ defined as above, this effect is captured by the regression coefficient for $z_1$ (as $z_2$ and $z_3$ are held constant).

### A1.3 Computation of the associated change in the mean outcome between any two given compositions

Consider two example compositions that were used for illustration purpose in the main text: namely, 10h (41.7%) sitting, 3h (12.5%) standing, 2h (8.3%) stepping, and 9h (37.5%) sleeping per day, and 7.6h (31.7%) sitting, 5.4h (22.5%) standing, 2h (8.3%), and 9h (37.5%) sleeping per day, denoted as $x_1$ and $x_2$, respectively. By using the perturbation and power operations defined in A1.1, the change from $x_1$ to $x_2$ equals $x_2 \ominus x_1$. The *ilr*-transformation of $x_2 \ominus x_1$ is $ilr(x_2) - ilr(x_1)$ because ilr-transformation is an isometric isomorphism between $S^4$ and $R^3$ (Pawlowsky-Glahn et al., 2015). If we denote $ilr(x_1)$ as $P_1$ and $ilr(x_2)$ as $P_2$ and the *ilr*-coordinates defined in the equation (3) in the main text are used, then $P_1$ and $P_2$ are just two points in $R^3$ with $P_1 = (-1.02, 0.53, -0.78)$ and $P_2 = (-1.12, 0.06, -0.34)$, and the change from $x_1$ to $x_2$ can be characterized by the change from $P_1$ to $P_2$ with difference of -0.1, -0.48, and 0.44 in each coordinate $z_1$, $z_2$, and $z_3$. With a fitted linear model $\hat{E}(Y) = \hat{\beta}_0 + \hat{\beta}_1 z_1 + \hat{\beta}_2 z_2 + \hat{\beta}_3 z_3$, the effect of time reallocation corresponding to the change from $x_1$ to $x_2$ is then $-0.1\hat{\beta}_1 - 0.48\hat{\beta}_2 + 0.44\hat{\beta}_3$.

## A2. Supplemental Methods and Materials for Latent Profile Analysis

### A2.1. Underlying probabilistic model in LPA

Suppose we observe three continuous variables $(X_1, X_2, X_3)$ and assume there are $K$ hidden profiles (call each profile $G_k$, $k = 1, \ldots, K$) underlying the observed data. LPA assumes the likelihood of any observation from the data follows

$$P(x_1, x_2, x_3) = \sum_{k=1}^{K} P(G_k) P(x_1, x_2, x_3 | G_k)$$

where $P(G_k)$ is the probability of observing latent profile $G_k$, k = 1, …. K; $P(x_1, x_2, x_3|G_k)$ is the probability density function for a multivariate normal distribution $N(\mu_k, \Sigma_k)$, k=1,…,K with ith profile specific mean vector $\mu_i$ and covariance matrix $\Sigma_k$. In this framework, $P(G_k)$, $\mu_k$, $\Sigma_k$ are all unknown quantities for each profile. If they are known or can be estimated from the data, we can then compute $P(G_k|x_1, x_2, x_3)$ the probability of an observation belonging to a latent profile $G_k$ (or called posterior probability) by using Bayes rule
$P(G_k|x_1, x_2, x_3) = \frac{P(G_k)P(x_1,x_2,x_3|G_k)}{\sum_{k=1}^{K} P(G_k)P(x_1,x_2,x_3|G_k)}$ for each profile k = 1,…,K.

The key components of LPA are to determine (1) the number of underlying latent profiles, and (2) variance-covariance structures (e.g. whether variances and covariances could vary between profiles) for the multivariate normal distributions, both of which can be tuned by using the fit statistics and other criteria we discussed in the main text.

Although it is generally assumed that mean vectors differ across profiles, LPA model can also vary in terms of how the class-specific (co)variance matrices of the indicator variables are constrained or allowed to vary within and between classes. The four commonly used models are listed below (Wardenaar 2021):

(1) The variances of individual profile indicators are set to be the same across classes but the covariances between them are set to be 0 (i.e. independent under the normal assumption)
(2) Both the variances and covariances of profile indicators are set to be the same across classes, but the covariances can be non-zero.
(3) The variances of profile indicators can differ between classes but the covariances are set to be 0.
(4) Both the variances and covariances can differ between classes and the covariances can be non-zero (most flexible model).

**A2.2. Latent Gold 6.0 Syntax for step 3 adjusted association analysis of latent profiles with CASI IRT score using improved BCH method to account for class uncertainty**

```
options
   missing  excludeall;
   output
      parameters=first  betaopts=wl standarderrors=robust profile probmeans=posterior
      estimatedvalues=model iterationdetails
   reorderclasses;
   step3 modal bch;
variables
   independent
    age_cat nominal,
    gender nominal,
    race_eth nominal,
    education numeric,
    bmi numeric,
    CESD_Score numeric,
    MED1 nominal;
   dependent casi_irt continuous;
   latent
    Cluster nominal coding = 3 posterior = (Clu#1, Clu#2, Clu#3, Clu#4);
equations
   casi_irt <- 1 + Cluster + age_cat + gender + race_eth + education + bmi + CESD_Score + MED1;
```

## A.4 Supplemental Tables

**Table S1. Variation matrix for the 24HAC composition of sit, stand, step and sleep (N=1034).[1]**

|       | Sit  | Stand | Step | Sleep |
|-------|------|-------|------|-------|
| Sit   | 0    | 0.43  | 0.54 | 0.08  |
| Stand | 0.43 | 0     | 0.23 | 0.27  |
| Step  | 0.54 | 0.23  | 0    | 0.42  |
| Sleep | 0.08 | 0.27  | 0.42 | 0     |

[1]Each entry is the standard deviation of the logarithm of the ratio between two activities in the sample.

**Table S2 Comparisons of the 24-hour Activity compositional mean by participant characteristics (N = 1034).** P-values correspond to the James multivariate analysis of variance test comparing the means of *ilr*-transformed data, assuming unequal variance between groups. The number of observations in subgroups may not sum to 1034 due to missing values.

|  | N (%) | Sitting, h/day (%) | Standing, h/day (%) | Stepping, h/day (%) | Sleeping, h/day (%) | p-value |
|---|---|---|---|---|---|---|
| *Overall* | 1034 (100%) | 10.2 (42.6%) | 3.7 (15.3%) | 1.2 (5.2%) | 8.8 (36.9%) | |
| *Gender* | | | | | | <0.001 |
| Male | 457 (44.2%) | 10.6 (44.2%) | 3.4 (14.2%) | 1.3 (5.3%) | 8.7 (36.3%) | |
| Female | 577 (55.8%) | 9.9 (41.2%) | 4.0 (16.5%) | 1.3 (5.2%) | 8.9 (37.1%) | |
| *Age* | | | | | | <0.001 |
| 65-74 | 433 (41.9%) | 10.0 (41.5%) | 3.9 (16.2%) | 1.5 (6.3%) | 8.6 (36%) | |
| 75-84 | 425 (41.1%) | 10.2 (42.4%) | 3.7 (15.5%) | 1.2 (5.2%) | 8.9 (37%) | |
| 85+ | 176 (17%) | 10.8 (45.1%) | 3.2 (13.5%) | 0.9 (3.6%) | 9.1 (37.8%) | |
| *Race/Ethnicity[1]* | | | | | | 0.002 |
| Non-Hispanic White | 920 (89%) | 10.2 (42.5%) | 3.7 (15.3%) | 1.3 (5.3%) | 8.9 (36.9%) | |
| Hispanic/Other race | 109 (0.5%) | 10.3 (42.8%) | 3.9 (16.4%) | 1.1 (4.7%) | 8.6 (36.1%) | |
| *Self-rated Health* | | | | | | <0.001 |
| Excellent/Very good/Good | 952 (92.1%) | 10.1 (42.2%) | 3.8 (15.7%) | 1.3 (5.5%) | 8.8 (36.5%) | |
| Fair/Poor | 82 (7.9%) | 11 (45.9%) | 2.8 (11.8%) | 0.7 (3.1%) | 9.4 (39.2%) | |
| *Physical function score[1]* | | | | | | <0.001 |
| <= 10 | 723 (69.9%) | 10.3 (42.9%) | 3.7 (15.3%) | 1.2 (5.1%) | 8.8 (36.7%) | |
| >10 | 243 (23.5%) | 9.7 (40.6%) | 4.0 (16.5%) | 1.6 (6.7%) | 8.7 (36.3%) | |
| *Ability to walk half mile[1]* | | | | | | <0.001 |
| No difficulty | 771 (74.6%) | 9.9 (41.1%) | 4 (16.5%) | 1.5 (6.2%) | 8.7 (36.2%) | |
| Some difficulty | 149 (14.4%) | 10.7 (44.4%) | 3.4 (14.1%) | 0.9 (3.9%) | 9 (37.6%) | |
| A lot difficulty/Unable | 110 (10.6%) | 11.6 (48.5%) | 2.5 (10.5%) | 0.6 (2.6%) | 9.2 (38.4%) | |
| *Sleep duration/day* | | | | | | <0.001 |
| <6 hrs | 17 (1.6%) | 12.3 (51.3%) | 4.5 (18.6%) | 1.3 (5.3%) | 5.9 (24.7%) | |
| 6-9 hrs | 669 (64.7%) | 10.4 (43.5%) | 3.9 (16.2%) | 1.4 (5.7%) | 8.3 (34.7%) | |
| 9+ hrs | 348 (33.7%) | 9.6 (40%) | 3.3 (13.8%) | 1.1 (4.5%) | 10 (41.7%) | |
| *Sleep quality[1]* | | | | | | 0.034 |
| Very poor to fair | 342 (33.1%) | 10.4 (43.2%) | 3.6 (15%) | 1.2 (5%) | 8.8 (36.8%) | |
| good to very good | 596 (57.6%) | 10.1 (41.9%) | 3.8 (15.9%) | 1.3 (5.6%) | 8.8 (36.6%) | |
| *CASI IRT score* | | | | | | <0.001 |
| <= 0 | 198 (19.1%) | 10.6 (44%) | 3.4 (14.3%) | 1.1 (4.4%) | 9 (37.3%) | |
| (0, 1] | 506 (48.9%) | 10.2 (42.4%) | 3.7 (15.5%) | 1.3 (5.3%) | 8.8 (36.8%) | |
| >1 | 330 (31.9%) | 10 (41.8%) | 3.8 (16%) | 1.4 (5.8%) | 8.7 (36.4%) | |

[1] Missing data: race/ethnicity N = 5; physical function N = 68; ability to walk half a mile N = 4; sleep quality N = 96.

**Table S3. Fit statistics for latent profile models with 2-6 profiles. Bold font numbers indicate the best results in terms of each fit statistic (N = 1034).[1]**

| Class Number | LogLik | AIC | BIC | CAIC | SABIC | ICL-BIC | N min (%) | LMR p-value | BLRT p-value | Entropy[2] |
|---|---|---|---|---|---|---|---|---|---|---|
| 2 | 5595.1 | -11152.3 | -11058.4 | -11039.4 | -11118.8 | **-10337.8** | 365 (35.3%) | | | 0.50 |
| 3 | 5648.8 | -11239.6 | **-11096.4** | **-11067.4** | -11188.5 | -10118.8 | 177 (17.1%) | <0.001 | <0.001 | 0.57 |
| 4 | 5673.5 | -11269.1 | -11076.4 | -11037.4 | **-11200.2** | -9784.0 | 132 (12.8%) | <0.001 | <0.001 | 0.55 |
| 5 | 5688.4 | -11278.8 | -11036.6 | -10987.6 | -11192.3 | -9550.4 | 80 (7.7%) | 0.022 | <0.001 | 0.55 |
| 6 | 5701.7 | **-11285.4** | -10993.9 | -10934.9 | -11181.3 | -9581.9 | 46 (4.4%) | 0.103 | <0.001 | 0.62 |

[1] Abbreviations: AIC (Akaike's Information Criterion), BIC (Bayesian Information Criteria), CAIC (Consistent Akaike's Information Criterion), SABIC (sample adjusted Bayesian Information Criteria), ICL-BIC (integrated complete likelihood-Bayesian Information Criteria), LMR (Lo-Mendell-Rubin likelihood ratio test), BLRT (Bootstrap likelihood ratio test).

[2] Entropy statistic ranges from 0 to 1 with higher values indicative of lower uncertainty in profile classification; 1 means perfect classification (i.e., without uncertainty) (Pastor et al., 2007). For example, when there are K latent classes, the formula to calculate this entropy based on N observations with each observation having probabilities $p_1, \ldots, p_K$ belonging to class 1 to class K, respectively, is $1 - \frac{\sum_{i=1}^{N}\sum_{k=1}^{K}(-p_{ik}\log(p_{ik}))}{N\log(K)}$.

**Table S4. Estimated profile-specific mean and variance-covariance matrix of the three profile indicators (sit, stand, step) in the final 4-class model (N = 1034).**

|  | Profile 1, most active (15.9%[1]) | Profile 2, moderately active low sleeper (24.4%[1]) | Profile 3, average activity (40.3%[1]) | Profile 4, least active (19.4%[1]) |
|---|---|---|---|---|
| **Mean (hrs/day)** | | | | |
| Sit | 7.6 | 9.6 | 10.3 | 12.1 |
| Stand | 5.8 | 4.7 | 3.5 | 2.4 |
| Step | 1.9 | 1.7 | 1.3 | 0.7 |
| Sleep[2] | 8.7 | 7.9 | 8.9 | 8.8 |
| **SD (hrs/day)** | | | | |
| Sit | 1.4 | 1.6 | 1.3 | 1.7 |
| Stand | 1.8 | 1.1 | 0.9 | 1.0 |
| Step | 0.8 | 0.5 | 0.4 | 0.3 |
| Sleep[2] | 1.1 | 1.0 | 0.6 | 1.4 |
| **Correlation** | | | | |
| Sit-Stand | -0.8 | -0.7 | -0.8 | -0.6 |
| Sit-Step | -0.1 | -0.3 | -0.5 | -0.4 |
| Stand-Step | -0.2 | -0.2 | 0.3 | 0.7 |
| Sit-Sleep[2] | <0.01 | -0.15 | -0.09 | -0.17 |
| Stand-Sleep[2] | -0.11 | 0.01 | 0.02 | -0.05 |
| Step-Sleep[2] | -0.07 | 0.05 | <0.01 | -0.06 |

[1] Estimated probability of each profile
[2] Sleep was not included in the LPA analyses. However, because the proportions of four activities sum to 1, the SD of sleep and its correlation with other activities can be derived. For example, the variance of sleep equals var(1-sit-stand-step) = var(sit + stand + step) = var(sit) + var(stand) + var(step) + 2cov(sit, stand) + 2 cov(sit, step) + 2 cov(stand, step).

**Table S5.** Expected misclassification errors for modal assignment based on the final 4-class LPA model presented in the main text (N = 1034).[1]

|  | Profile according to modal assignment | | | |
|---|---|---|---|---|
| **Latent profile** | 1 | 2 | 3 | 4 |
| 1 | 0.797 | 0.103 | 0.089 | 0.012 |
| 2 | 0.215 | 0.669 | 0.017 | 0.010 |
| 3 | 0.216 | 0.036 | 0.747 | 0.001 |
| 4 | 0.142 | 0.233 | 0.007 | 0.618 |

[1] Each entry is the estimated probability that an observation belonging to a latent profile (indicated by row number) is assigned to each profile (indicated by column number).

**A.5 Supplemental Figures**

**Figure S1. Scatterplots of data points in the simplex space for the 3-part composition of sit, stand, and step (left graph), and in 2D real space with isometric logratio coordinates $z_1$ and $z_2$. $z_1 = \sqrt{\frac{2}{3}} \ln \frac{Sit}{(Stand \times Step)}$ and $z_2 = \sqrt{\frac{1}{2}} \ln \frac{Stand}{Step}$ (right graph).** The red dot on each graph below is an example data point located in the simplex space (before *ilr*-transformation) and 2D real space (after *ilr*-transformation), respectively.

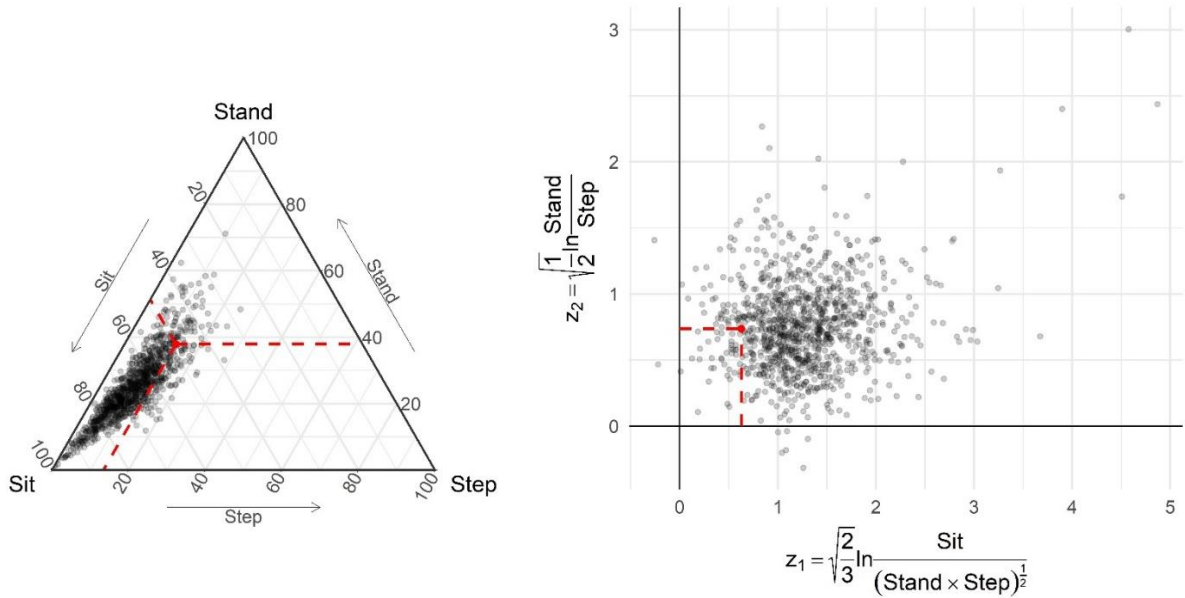

**Figure S2. Distributions of percentage time per day spent in each behavior (N=1034).** Upper triangular entries are Pearson's correlation coefficients. Superscript stars indicate the level of significance i.e. *** indicates p-value < 0.001.

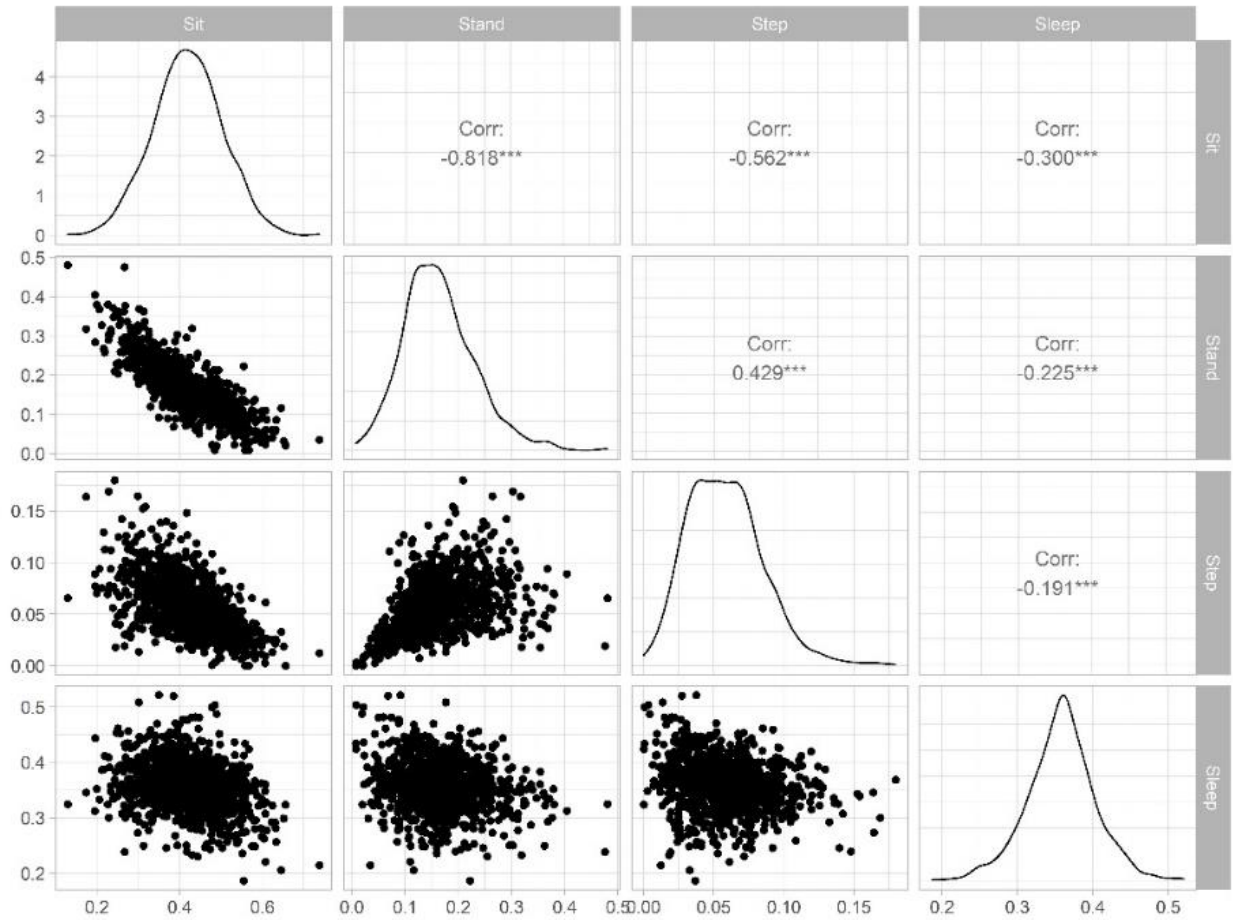

**Figure S3. Distributions of *ilr*-transformed data with ilr-coordinates $z_1, z_2, z_3$ specified in equation (2) in the main manuscript, Section 3.2 (N=1034).** Upper triangular entries are Pearson's correlation coefficients. Superscript stars indicate the level of significance i.e. *** indicates p-value < 0.001; ** indicates p-value <0.01.

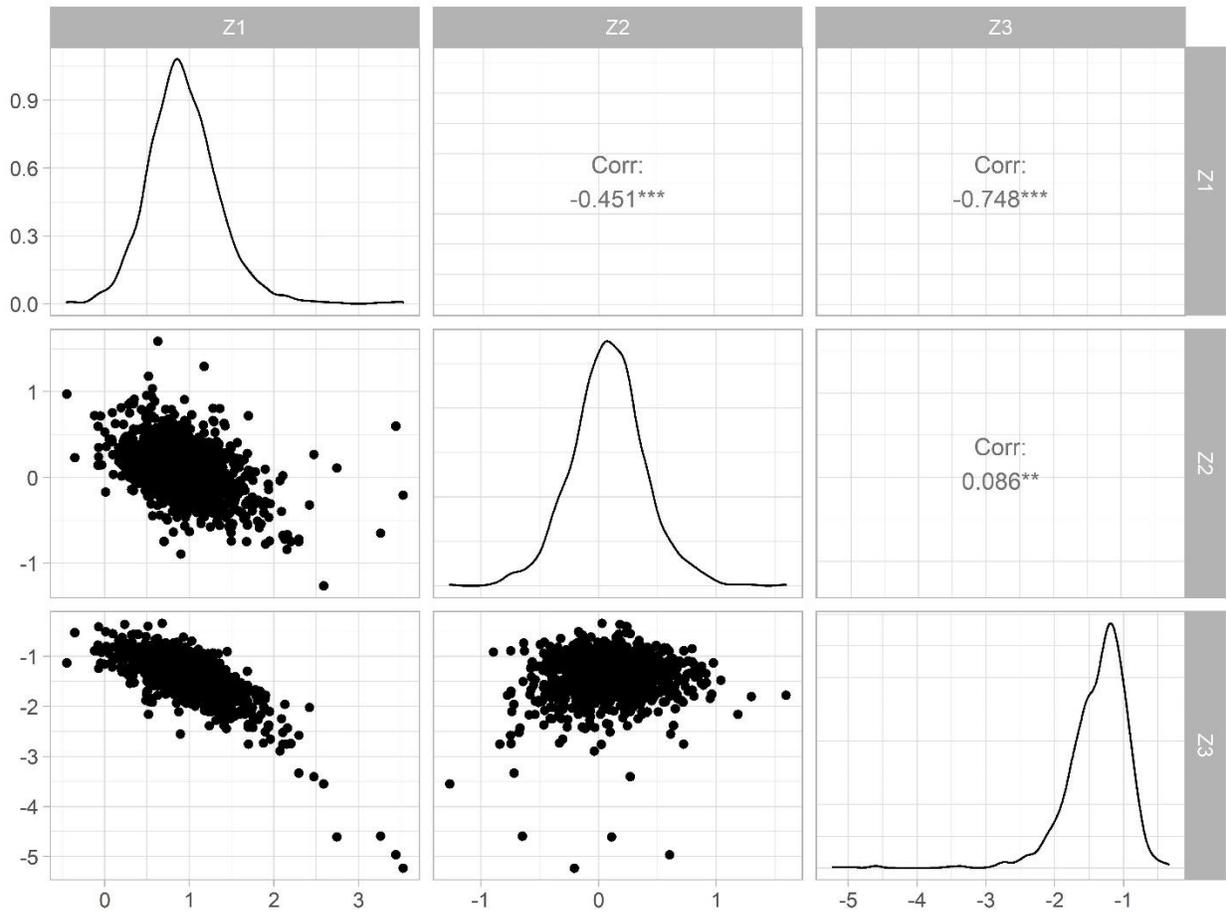

**Figure S4. Fitted 24HAC profiles from latent profile analysis (3-class solution), where each individual is assigned the class with maximum probability. The boxplots present sample quartiles (N=1034).** The number (%) of subjects in profile 1-3 is, respectively, 177 (17.1%), 622 (60.2%), and 235 (22.7%).

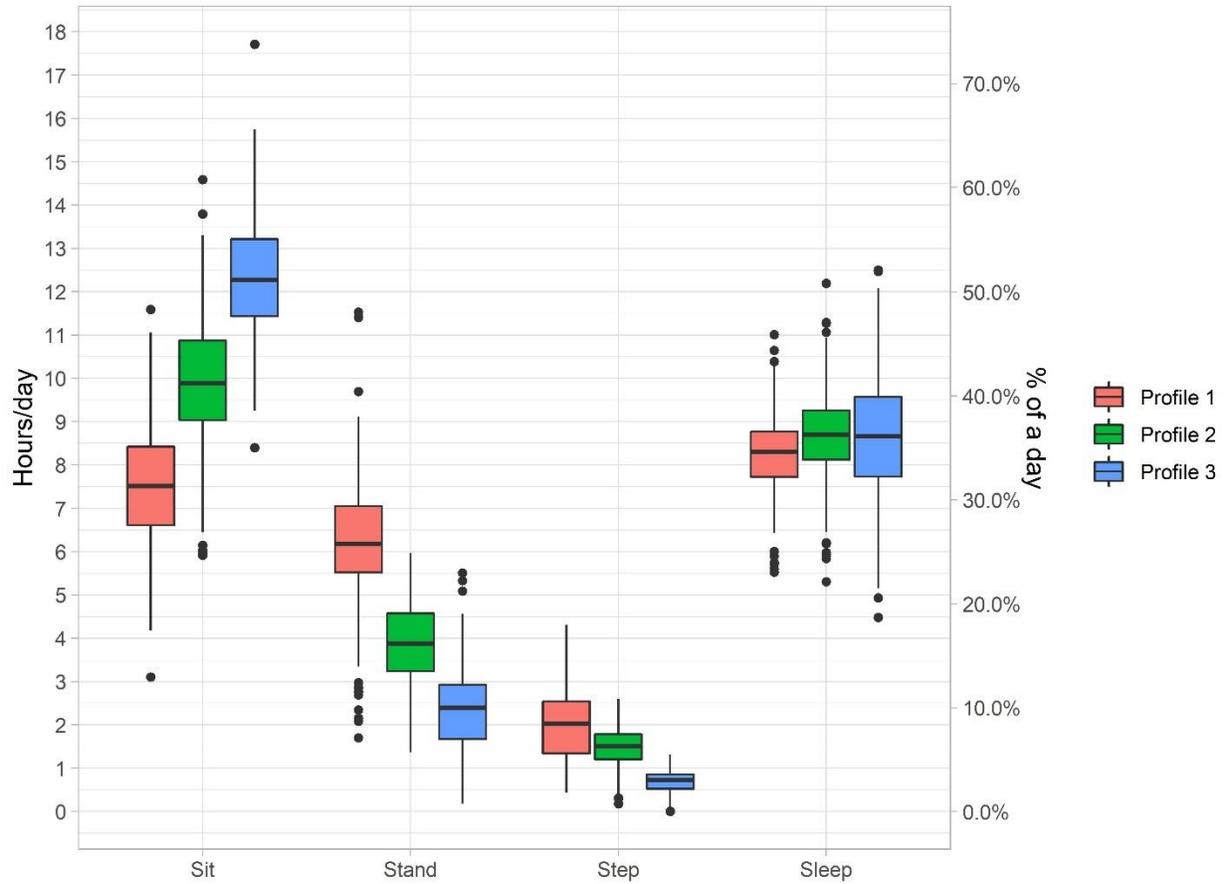